\documentclass[twocolumn]{aastex61}
\usepackage{amsmath,amstext}
\let\oldAA\AA
\renewcommand{\AA}{\text{\normalfont\oldAA}}
\usepackage{graphicx}
\usepackage[english]{babel}
\usepackage[T1]{fontenc}
\usepackage{comment}
\usepackage{apjfonts} 
\usepackage[figure,figure*]{hypcap}
\usepackage{epstopdf}
\usepackage{natbib}

\newcommand{\ergs}	{\ifmmode {\rm erg\,s}^{-1} \else erg s$^{-1}$\fi}

\newcommand{\civ}{\ifmmode {\rm C}\,\textsc{iv}  \else C\,\textsc{iv}\fi}
\newcommand{\CIV}{\ifmmode {\rm C}\,\textsc{iv}\,\lambda1549 \else C\,\textsc{iv}\,$\lambda1549$\fi}
\newcommand{\mgii}{\ifmmode {\rm Mg}\,\textsc{ii} \else Mg\,\textsc{ii}\fi}
\newcommand{\MgII}{\ifmmode {\rm Mg}\,\textsc{ii}\,\lambda2798 \else Mg\,\textsc{ii}\,$\lambda2798$\fi}
\defcitealias{Cap15}{C15}

\begin{document}

\title{AGN evolution from galaxy evolution viewpoint - II}
\shorttitle{AGN evolution from galaxy evolution viewpoint - II}
\shortauthors{Caplar, Lilly \& Trakhtenbrot}
\author{Neven Caplar}
\affiliation{Department of Physics, ETH Zurich, Wolfgang-Pauli-Strasse 27, 8093, Zurich, Switzerland}
\affiliation{Department of Astrophysical Sciences, Princeton University, 4 Ivy Ln.,  Princeton, NJ 08544, USA}
 \email{ncaplar@princeton.edu}

\author{Simon J. Lilly }
\affiliation{Department of Physics, ETH Zurich, Wolfgang-Pauli-Strasse 27, 8093, Zurich, Switzerland}

\author{Benny Trakhtenbrot}
\affiliation{Department of Physics, ETH Zurich, Wolfgang-Pauli-Strasse 27, 8093, Zurich, Switzerland}
\affiliation{School of Physics and Astronomy, Tel Aviv University, Tel Aviv 69978, Israel}
\begin{abstract}
In order to relate the observed evolution of the galaxy stellar mass function and the luminosity function of active galactic nuclei (AGN),  we explore a co-evolution scenario in which AGN are associated only with the very last phases of the star-forming life of a galaxy. We derive analytically the connections between the parameters of the observed quasar luminosity functions and galaxy mass functions. 
The $(m_{\rm bh}/m_{*})_{Qing}$ associated with quenching is given by the ratio of the global black hole accretion rate density (BHARD) and star-formation rate density (SFRD) at the epoch in question. Observational data on the SFRD and BHARD suggests $(m_{\rm bh}/m_{*})_{Qing} \propto (1+z)^{1.5}$ below redshift 2.
This evolution reproduces the observed mass-luminosity plane of SDSS quasars, and also reproduces the local $m_{\rm bh}/m_{*}$ relation in passive galaxies.  
The characteristic Eddington ratio, $\lambda^*$, is derived from both the BHARD/SFRD ratio and the evolving $L^*$ of the AGN population.  
This increases up to $z \sim 2$ as $\lambda^* \propto (1+z)^{2.5}$ but at higher redshifts, $\lambda^*$ stabilizes at the physically interesting Eddington limit, $\lambda^* \sim 1$.   The new model may be thought of as an opposite extreme to our earlier co-evolution scenario in \cite{Cap15}. The main observable difference between the two co-evolution scenarios, presented here and in \cite{Cap15}, is in the active fraction of low mass star-forming galaxies.  
We compare the predictions with the data from deep multi-wavelength surveys and find that the ``quenching'' scenario developed in the current paper is much to be preferred.
\end{abstract}

\setwatermarkfontsize{3in} 


\section{Introduction}

Recently there has been significant interest in the co-evolution of galaxies and their supermassive black holes and the possible connections between these two populations. This has been primarily motivated by the observed correlations between the masses of the central black holes and the properties of host galaxies \citep{Fer00,Geb00,Mar03,Har04,San11,Mar13,Kor13,Rei15}. Additionally, there are qualitative similarities between the cosmic evolution of star formation rate density (SFRD) and that of the luminosity density of active galactic nuclei (AGN) which tracks the cosmic black hole accretion rate density (BHARD). Both of these functions rise with increasing redshift up to $z\simeq2$ where they reach a broad maximum before falling again at higher redshifts \citep{Boy98, Hec04, Has05, Sil09, Mad14,Air15}.

Several recent studies have demonstrated the strength of a phenomenological approach to understanding the evolving population of AGN and their host galaxies. 
\cite{Wei17} showed that the population of accreting black holes in the local universe, and specifically their luminosity functions, can be reproduced by two main AGN populations, with markedly different (mass-independent) Eddington ratio distribution functions: 
radiatively efficient AGN hosted in optically blue and green galaxies, and 
radiatively inefficient AGN hosted in red galaxies, ``quenched'' galaxies.
\cite{Air18} used a combination of extragalactic fields with deep X-ray coverage to probe the dependence of the Eddington ratio distribution function on host mass (and on redshift), finding that the probability for a galaxy to host an AGN depends on the galaxy mass. This is similar to the conclusion of \cite{Ber18}, who have argued that mass dependence is needed in order to reproduce the flat relation between host SFR and the AGN (X-ray) luminosity. 
\cite{Geo17} has emphasized the need for a strong redshift evolution of the AGN duty cycle, using similar data to that used by \cite{Air18}.

In \cite{Cap15} (hereafter \citetalias{Cap15}) we explored a scenario for the co-evolution of galaxies and their central black holes through comparisons of the evolving mass function of star-forming galaxies with the evolving X-ray luminosity function of AGN.  
In \citetalias{Cap15} we constructed a simple ``convolution model'' that enabled us to link the redshift-dependent parameters describing the AGN luminosity function and the galaxy mass function.   We assumed in that paper that black hole growth occurred on similar time scales as the build up of the stellar mass, i.e. that there was a steady build-up of the black hole mass in a galaxy as the stellar mass increased, thereby producing a characteristic ratio between stellar and black hole masses in galaxies.   We allowed that this characteristic ratio might however itself evolve slowly with epoch as the overall population evolved.   We showed that, in such a scenario, there are simple analytic relations between the parameters describing the Schechter galaxy mass function, i.e., M$^*$, $\phi_{SF}^*$ and $\alpha$, and the parameters that are used to describe the double power-law of the quasar luminosity function (QLF), i.e., L$^*$, $\phi_{QLF}^*$ and $\gamma_1$ and $\gamma_2$.  These relations are a consequence of the fact that QLF is a convolution of the black hole mass function and Eddington ratio function, the former being itself a convolution of the galaxy mass function and the relation between the galaxy and black hole mass, which we assumed to be a Gaussian around some fiducial (but possibly evolving) value. Within the framework of this model, our observational knowledge about the evolution of the galaxy mass function and the QLF was used to constrain the allowed evolution of the characteristic Eddington ratio and the evolving galaxy-black hole relation $m_{\rm bh}/m_{*}$.  A short summary of this earlier paper and its methodology is given in Section \ref{sec:Reprise}. \par

In \citetalias{Cap15} we found that significant evolution in the $m_{\rm bh}/m_{*}$ ratio was preferred to the scenario in which $m_{\rm bh}/m_{*}$ stays constant with redshift. This was based on the comparison of the models which used different $m_{\rm bh}/m_{*}$ evolution with the distribution of quasars in the BH mass-luminosity plane using SDSS data from \cite{She11} and \cite{Tra12}. In order to explain both the shape of the QLF and the distribution in the mass-luminosity plane at different redshifts, we found that we had to impose an evolution of something like $m_{\rm bh}/m_{*} \propto (1+z)^{n}$ with $n \sim 2$.  This was strongly preferred over the the non-evolving case of $n=0$.  

Whether such evolution of the mass-scaling between black holes and their host galaxies occurs has been hotly debated,  with a number of studies suggesting redshift evolution, with black holes being more massive, respective to host galaxy mass, at higher redshift  (\citealp{Pen06}; \citealp{Dec10}; \citealp{Mer10}; \citealp{Tra10}; \citealp{Ben11}; \citealp{Sij14}; \citealp{Tra15}). On the other hand, some other studies have claimed to see no significant evolution (\citealp{Jah09}; \citealp{Cis11}; \citealp{Mul12}; \citealp{Schr13}) or that all of the evolution found in the observational studies is due to selection effects (\citealp{Lau07}; \citealp{SchuWis11}; \citealp{Schu14}). \par

Recently, observational studies in the local Universe have provided valuable insights on this question. \cite{Fer15} and \cite{Fer17} presented evidence for the existence of very compact quenched galaxies in the local Universe that have much larger black hole masses than could be expected from the scaling relations. Given the compactness of these galaxies and their old mean mass-weighted ages, they can be interpreted as relics of a typical population that quenched at $z \sim 2$. The fact that their central black holes are more massive than seen generally in the local Universe supports the idea that the $m_{\rm bh}/m_{*}$ ratio was larger at $z\sim 2 $ than at lower redshifts. The same conclusion was reached by \cite{Bar16} using the result from the Eagle simulation \citep{Sch15Eagle,Cra15}. Different lines of evidence for the mass ratio evolution comes from \cite{Rei15} who surveyed the $m_{\rm bh}/m_{*}$ in the active systems in the nearby Universe ($z\lesssim 0.06$) and found the $m_{\rm bh}/m_{*}$ ratio in these systems to be more than an order of magnitude lower than the $m_{\rm bh}/m_{*}$ ratio found in the quenched systems. 
Such an offset is consistent with the scenario where the $m_{\rm bh}/m_{*}$ ratio is decreasing with cosmic epoch, as quenched galaxies would maintain a mass ratio representative of the active systems at the time of their quenching. 
This would naturally lead to quenched galaxies having larger $m_{\rm bh}/m_{*}$ than locally observed active galaxies. 
All of these approaches described above imply the evolving $m_{\rm bh}/m_{*}$ relation, as suggested in \citetalias{Cap15} through quite independent arguments. \par

However, there was an obvious weakness in the \citetalias{Cap15} analysis. This was that it did not explicitly include information on the star-formation history of galaxies in the Universe and therefore did not ``close the loop'' between the global build-up of BH mass, the build-up of stellar mass and the evolving $m_{\rm bh}/m_{*}$ ratio. In fact, a possible tension in that model was already apparent in \citetalias{Cap15}, in that it was hard to get a fast enough decline with cosmic epoch in the $m_{\rm bh}/m_{*}$ ratio, even if there was no BH growth at all. In this paper we aim to expand and develop the analysis from \citetalias{Cap15} by building a more self-consistent model in which we include information about the mass build-up histories of both the galaxy and black hole population. This was not the case in \citetalias{Cap15} - all of our conclusions about the evolution of quantities, such as the black hole - galaxy mass ratio and the evolution of the normalization of the QLF and galaxy mass functions, were based on observational data but were not connected with quantities describing growth of stellar and black hole population.  For example, in the previous work we have put in the mass ratio evolution ``by hand'' to explain the observed mass ratio evolution in the mass-luminosity plane i.e., the mass ratio evolution was an {\it input} for the model that was adjusted to produce consistency with the mass-luminosity plane. In the work described here, we will use the cosmic SFRD and BHARD in order to follow the mass build up of both population over time and the form of the mass-luminosity plane will now be an {\it output} prediction of the model that can be used to check validity of our assumptions. 
\par

We will find that in order to simultaneously satisfy the observational constraints from the SFRD and BHARD, we will need to implement a major change in the underlying scenario for BH growth in galaxies. Instead of the scenario explored in \citetalias{Cap15} where both galaxies and their central black holes grew ``together'' over the whole star-forming life of a galaxy, we are forced towards the other extreme in which most of the black hole growth is associated with a short growth phase only at the end of the star-forming life of the galaxy.  In other words, we will have the black hole growth associated in some way with the ''quenching'' of star-formation in galaxies.  

\par
We will use the quenching formalism of \cite{Pen10}.  Specifically we will consider the ``mass-quenching'' of \cite{Pen10} and ignore any ``environmental-quenching''.  The primary reason for this is that it is mass-quenching which is most directly tied to the evolution of the mass function of star-forming galaxies, since mass-independent environment-quenching does not change the shape of the mass function.  Ignoring environment-quenching may or may not be justified physically, but it is anyway sub-dominant at the galaxy masses of most interest here.
An important insight from \cite{Pen10} was that the specific star formation rate (sSFR) of the Main Sequence of galaxies controls not only the growth of the stellar mass of galaxies but also the rate of (mass-) quenching in the population and therefore, in the new scenario considered in this paper, the growth of black holes. By directly linking both black hole growth and stellar mass growth to the Main Sequence sSFR in this way, we can effectively eliminate a variable from the analysis.  This enables us to break the degeneracy between the black hole masses and Eddington ratios that we encountered in \citetalias{Cap15}.  Whereas in that paper, this degeneracy could be broken only by inputing additional observational data in the form of the distribution of objects in the $(m_{\rm bh},L)$ plane, we will find in this paper that this distribution can be an output prediction of the model.   \par
Additional output from the model are the connections between galaxies and their respective black holes. Number of observational studies have tried to directly observe the connection between AGN activity, AGN mass, star-formation and quenching \citep{Lut10, Sha10, Har12, Mul12, Ros12, Sta15, Sta17, Lan17}. These studies have shown that the star formation of AGN are in quantitative agreement with the population of the star-forming galaxies with the same stellar mass. Directly inferring the connection between star-formation and AGN activity is difficult without appropriate modeling as the correlation can be masked  due to different time scales of AGN and galaxy processes \cite[e.g.,][]{Hic14,Vol15,Ber18}. We will investigate these questions in context of our model in Section \ref{sec:Dif}.

The layout of the paper is as follows. In Section \ref{sec:Reprise} we provide an reprise of \citetalias{Cap15} that may be skipped by readers familiar with that paper. We also point out the tension between the co-evolution scenario that was basis of the \citetalias{Cap15} analysis, the observed mass growth rates for black holes and galaxies, and the inferred black hole-galaxy mass evolution. In Section \ref{sec:NewmodelQ} we develop the quenching scenario that is the basis of this paper. This scenario naturally resolves the tension because the observed mass black hole-galaxy mass ratio responds quickly to changes in the mass accretion rate. We also derive analytically connections between parameters describing observed QLF and black hole mass function and the cosmic mass growth density for galaxies and black holes. In Section \ref{sec:Determination} we determine the evolution of the black hole - galaxy mass ratio, the normalization of the AGN mass function, and the characteristic Eddington ratio using analytical connections from previous section and using observational constraints. We find that black hole - galaxy mass ratio evolves as $(1+z)^{1.5}$ up to redshift 2, quite similar to $(1+z)^{2}$ we suggested in \citetalias{Cap15}. We also find that inferred evolution of the normalization and the Eddington ratio is very similar to the evolution suggested in \citetalias{Cap15}. In section \ref{sec:Quant} we show results of our quantitative analysis in which we compare results of this scenario directly with observations of active AGN in mass-luminosity plane and with observations of quiescent black holes in local Universe. We show that the new quenching scenario reproduces these observations in a similar fashion as the co-evolution scenario from \citetalias{Cap15}. This is a consequence of the similar evolution of the black hole-galaxy mass ratio, the normalization and  the Eddington ratio. In Section \ref{sec:Dif} we discuss observation which are able to differentiate these two scenarios. We compare predictions from both scenarios for deep x-ray and optical observations in COSMOS and CDF-S fields and find that the quenching scenario presented here provides better description to the data. We summarize in Section \ref{sec:Sum}. \par

Throughout the Paper, we will assume a $\Lambda$CDM cosmology, with parameters $\Omega_{M}=0.3$, $\Omega_{\Lambda}=0.7$ and $H_{0}= 70$ km s$^{-1}$ Mpc$^{-1}$. 
AGN luminosities will be given in units of erg s$^{-1}$ and will refer to bolometric luminosities, unless specified otherwise. 
We use the term "dex" to denote the antilogarithm, i.e., $n$ dex = 10$^{n}$. 
We also define all distribution functions, i.e., the star-forming galaxy mass function $\phi_{SF}(m_{*})$, the associated star-forming galaxy black hole mass function $\phi_{bh}(m_{\rm bh})$, the mass function of galaxies which are undergoing mass quenching $\phi_{Qing}(m_{*})$, the AGN mass function $\phi_{bh}(m_{AGN})$, the AGN luminosity function $\phi(L)$, and the probability distribution of Eddington ratio $\xi(\lambda)$, in log$_{10}$ space. This leads to power-law exponents that differ by unity relative to distribution functions defined in linear space. 
The units of $\phi_{SF}(m_{*})$, $\phi_{bh}(m_{\rm bh})$, $\phi_{Qing}(m_{*})$, $\phi_{bh}(m_{AGN})$, and $\phi(L)$ are Mpc$^{-3}$ dex$^{-1}$. 
An overview of the parameters used throughout this paper is given in Table~ \ref{tab:parameters}.

\section{Reprise of Caplar et al. (2015)} \label{sec:Reprise}

\begin{figure*}[htp]
    \centering
    \includegraphics[width=.89\textwidth]{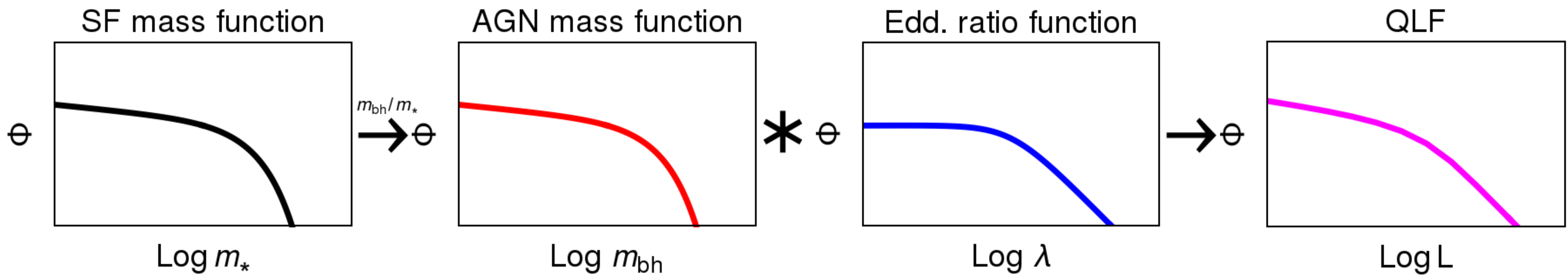}
    \caption{Sketch of the convolution model used in \citetalias{Cap15}. The quasar luminosity function is given by the convolution of the AGN black hole mass function and the Eddington ratio distribution. The AGN mass function, describing the mass distribution of actively acreting black holes, is derived by convolving the galaxy mass function of star-forming galaxies with an $m_{\rm bh}/m_{*}$ distribution. In this model all of the star-forming galaxies have the same chance to host AGN at a given moment in time.}
    \label{fig:OldModel}
\end{figure*}

The \citetalias{Cap15} analysis created a global model which used our knowledge about the evolving galaxy mass function as an input to try to explain the main features of the evolving QLF and thereby to empirically connect the galaxy and AGN populations. We show the main features of this concept in Figure \ref{fig:OldModel}. The quasar luminosity function is given by a convolution of the black hole mass function and the Eddington ratio function,
\begin{equation} \label{eq:CreateQLF}
\phi (L,z) = \int \phi_{bh} (m_{\rm bh},z) \xi (\lambda,z) d \log\lambda,
\end{equation}
where $\phi (L,z)$ is the QLF,  $\phi_{bh} (m_{\rm bh},z)$ is the black hole mass function and $\xi (\lambda,z)$ is the Eddington ratio distribution. 
In this expression, the redshift evolution of the QLF is well known, and the black hole mass function can be constructed by convolving the galaxy mass function of galaxies hosting an AGN (assumed to be the star-forming population) with a gaussian function describing the distribution of $m_{\rm bh}/m_{*}$. \par

For simplicity, we assumed that this black hole-galaxy mass relation is linear, and that all star-forming galaxies are equally like to host an active black hole. 
These assumptions correspond to a ``co-evolution'' scenario in which AGN are simply co-existing with their host galaxies with both the stellar mass and black hole mass growing more or less in step, with a (possible) slow evolution in th mean ratio over cosmic time. \par
The main input to this analysis is the cosmological evolution of the star-forming galaxy mass function. 
It is well described with a Schechter function, that is
\begin{equation} \label{eq:SchSF}
\phi_{SF} (m_{*}) \equiv \frac{d N}{ d \log m_{*}} =  \phi^{*}_{SF} \left(\frac{m_{*}}{M^{*}} \right)^{\alpha_{SF}}  \exp \left(-\frac{m_*}{M^{*}} \right),
\end{equation}
where $\phi^{*}_{SF}$ is the normalization, $\alpha_{SF}$ is the low mass slope and $M^{*}$ is the Schechter mass. 
It is known observationally (\citealp{Ilb13};  \citealp{Muz13}; \citealp{Tom14}; \citealp{Mor15}; ) that $M^{*}$ stays essentially constant with redshift up until at least redshift 3, especially if the faint end slope $\alpha$ is fixed, while $\phi^{*}_{SF}$ steadily falls towards higher redshift.  
Both the qualitative and quantitative evolution of these parameters is in accordance with the phenomenological model of galaxy growth and quenching by \cite{Pen10}.  \par

We then used the linear correlation between the galaxy and the black hole mass to construct the black hole mass function for black holes hosted in star-forming galaxies, which again has a broadly Schechter form, broadened by the dispersion in $M^{*}_{bh} = M^{*} \cdot m_{\rm bh}/m_{*}$ relation (sometimes called "modified Schechter function", see \cite{SchuWis10}). \par

The QLF is often described with the following broken power law function:
\begin{equation} \label{eq:QLF}
\Phi(L)\equiv \frac{d N}{d \mbox{log} L} = \frac{\phi^{*}_{QLF}}{(L/L^{*})^{\gamma_{1}} +(L/L^{*})^{\gamma_{2}}},
\end{equation}
where  $\phi^{*}_{QLF}$ is the normalization, $L^{*}$ is the break luminosity and $\gamma_{1}$ and $\gamma_{2}$ are, respectively, the faint and bright end slopes. From the shape of this function, the shape of the black hole mass function and the Equation \eqref{eq:CreateQLF} it follows that the Eddington ratio distribution function must also have broken power law shape:

\[
 \xi(\lambda) \equiv \frac{dN}{N d \log \lambda}= 
  \begin{cases} \label{eq:Ed}
                                   \frac{\xi^{*}_{\lambda}}{(\lambda/\lambda^{*})^{\delta_{1}}  +(\lambda/\lambda^{*})^{\delta_{2}} },  & \lambda>\lambda_{min} \\
                                    0 & \lambda<\lambda_{min} 
  \end{cases}
\]

where $\xi^{*}_{\lambda}$ is the normalization, $\lambda^{*}$ is characteristic Eddington ratio, $\delta_1$ and $\delta_2$ are, respectively, the low and high end slopes and $\lambda_{min}$ is the lower cutoff of the distribution. \par

We found that this simple model can fully explain the observed QLF. Due to the simplicity of the model, it is straightforward to directly and analytically connect the parameters describing the galaxy and AGN populations. 
Here we reproduce only the main equations which describe the normalization and the location of the characteristic luminosity, or ``knee'', of the double power-law QLF:

\begin{equation} \begin{split}\label{eq:ap8}
L^* & \cong 10^{38.1} \cdot \lambda^{*} \cdot M^{*}  \cdot \frac{m_{\rm bh}}{m_{*}} , \\
\phi^{*}_{QLF}& \cong \xi^{*}_{\lambda} \cdot \phi^{*}_{SF}.  
\end{split}\end{equation}

The first equation relates the characteristic quantities in Equation \eqref{eq:CreateQLF} and states that the characteristic $L^*$ luminosity of the QLF will be given (to within factors of order unity) by the product of the characteristic galaxy mass $M^*$, the characteristic Eddington ratio $\lambda^*$, and the average black-hole galaxy mass ratio. The second relation states that the number of actively accreting black holes, $\phi^{*}_{QLF}$, will be given by the product of the number density of star-forming galaxies, $\phi^{*}_{SF}$ and the normalisation of the knee of the Eddington ratio distribution $\xi^{*}_{\lambda}$, which is a general ``duty cycle'' of the black hole population, i.e., the probability that they are active. 

We then pointed out two interesting features of the redshift evolution of these parameters ($L^{*}$ and $\phi^{*}_{QLF}$). Firstly, the observed normalizations of both the star-forming galaxy mass function and the QLF evolve in a very similar manner to at least $z\simeq 3$ and possibly beyond.  As described above, in the model these two normalizations are directly connected via the normalization of the Eddington ratio distribution. The fact that the ratio of these two observed quantities (mass function and QLF) is constant implies that the normalization of the Eddington ratio function also stays constant with redshift, i.e., that the ``duty cycle'' of the AGN population is essentially constant with redshift. 
\par 

Second, the QLF $L^{*}$ changes significantly with redshift, with this evolution being roughly $L^{*} \propto (1+z)^{4}$ to $z\sim2$.
The luminosity corresponding to the ``break'' of the QLF, $L^{*}$, is directly connected with the value of the Schechter break in galaxy mass function, $M^{*}$, the break in the Eddington ratio distribution, $\lambda^{*}$, and the scaling ratio between galaxy and black hole mass, $m_{\rm bh}/m_{*}$. Given that there is strong evidence that $M^{*}$ does not change, the observed evolution in $L^{*}$ suggests evolution in the knee of the Eddington ratio distribution, or evolution in black hole - galaxy mass ratio or some combination of these two quantities.
\par

We could in principle break this degeneracy by modeling the distribution of AGN in the mass-luminosity plane, as observed in SDSS, including all the relevant selection cuts.   We found that the observed distribution could best be explained by splitting the evolution roughly equally between contributions from a $(1+z)^{2}$ change in $m_{\rm bh}/m_{*}$ and a $(1+z)^{2}$ increase in the break of the Eddington ratio distribution, $\lambda^{*}$.
We argued that this type of change is consistent with the observed $m_{\rm bh}/m_{*}$ relations in both star-forming and quenched galaxies in the local Universe and with the observations of $m_{\rm bh}/m_{*}$ ratio in active galaxies at higher redshifts. We also pointed out that this kind of evolution, when coupled with observed galaxy size evolution, naturally leads to a redshift \emph{independent} $m_{\rm bh}-\sigma_*$ relation. \par 

However, already in \citetalias{Cap15}, we pointed out that there was a shortcoming of the model.  Given that all galaxies were assumed to have roughly the same $m_{\rm bh}/m_{*}$ at a given redshift, we showed that it was very difficult to get this ratio to evolve fast enough at low redshifts ($z \lesssim 0.7$) simply because there is not enough star-formation to decrease $m_{\rm bh}/m_{*}$ even if there is no black hole growth at all.  We describe and elaborate on this problem further below. \par
For the galaxies on the Main Sequence, we can describe the expected increase of their stellar mass as
\begin{equation} \label{eq:rsSFR}
rsSFR (z)=\frac{\dot{m}_{*}}{m_{*}} =  \left(  \frac{-1}{H_{0}(1+z)\sqrt{\Omega_{M}(1+z)^{3}+\Omega_{\Lambda}}} \right)^{-1}  \frac{1}{m_{*}} \frac{d m_{*}}{dz},
\end{equation}
where $rsSFR$ is the ``reduced specific star-formation rate'' (see \citealp{Lil13}), $rsSFR = (1-R)sSFR$, where $R$ is the fraction of stellar mass that is returned during star formation, $R \sim 0.4$. If we use the expression for the redshift evolution of the Main Sequence \citep{Lil13}
\begin{equation} \label{eq:lillySFR}
rsSFR|_{ms} (z) = 0.07 (1+z)^{3} \mbox{ Gyr}^{-1},
\end{equation}
we find that these galaxies change their mass by $\sim 0.5$ dex between redshift 0 and redshift 0.7. This value is therefore a maximal change of the  $m_{\rm bh}/m_{*}$ ratio which is possible in the Universe. This is consistent with $(1+z)^{2}$ evolution of the $m_{\rm bh}/m_{*}$ ratio only if there is no black hole growth whatsoever during this period. At lower redshifts, the galaxy growth becomes even slower and as such it is not possible to create such strong evolution in the $m_{\rm bh}/m_{*}$ ratio. \par

Of course, we know that the case in which black holes do not accrete at all is not realistic. We can therefore estimate how the mass doubling time for the black holes would need to change if the $m_{\rm bh}/m_{*}$ ratio changes by the factor of ten from $z\sim0$ to $z\sim2$ -- the change implied by $m_{\rm bh}/m_{*} \propto (1+z)^{2}$ evolution. 
We use an equation describing mass growth of actively accreting BHs (i.e., AGN), which takes a similar shape to Eq.~\eqref{eq:rsSFR} :  
\begin{equation}\begin{split}
\label{eq:rBHAR}
&\frac{10^{38.1}}{10^{37.75}} \frac{1-\epsilon}{\epsilon}\left\langle \lambda \right\rangle  =  \frac{\dot{m}_{bh}}{m_{\rm bh}}\\& = \left(  \frac{-1}{H_{0}(1+z)\sqrt{\Omega_{M}(1+z)^{3}+\Omega_{\Lambda}}} \right)^{-1}  \frac{1}{m_{\rm bh}} \frac{d m_{\rm bh}}{dz}, 
\end{split}\end{equation}
where $\left\langle \lambda \right\rangle$ is the mean Eddington ratio of the population\footnote{The pre-factor $10^{38.1}$ in the nominator of Eq.~\ref{eq:rBHAR}, and elsewhere in this paper, comes from the definition of Eddington ratio, $L = 10^{38.1} m_{\rm bh} \lambda$, where the luminosity is given in erg s$^{-1}$ and black hole mass is given in $M_{\odot}$. 
In the denominator, the pre-factor $10^{37.75}$ comes from unit conversion in the expression $L=\dot{m}_{\rm bh} \epsilon/(1-\epsilon)$ from kg s$^{-1}$ to M$_{\odot}$ Gyr$^{-1}$ on the right hand side of Eq.~\ref{eq:rBHAR}, and again expressing luminosity in erg s$^{-1}$. 
With this conversion, the units in which Equation \eqref{eq:rBHAR} is expressed is Gyr$^{-1}$, which is same as in equivalent relations \eqref{eq:rsSFR} and \eqref{eq:lillySFR}. We have also suppressed explicitly writing $c$, the speed of light, in this and in the following equations throughout the paper.}. 
If we assume that both the galaxy and black hole populations had the same mass doubling times at $z\sim 2$ and there is no redshift evolution in the efficiency of BH accretion, we find that the mean Eddington ratio has to evolve as $\left\langle \lambda \right\rangle \propto (1+z)^{5.5}$ to create a change in the typical galaxy - black hole ratio by the factor of 10! 
This means that the black hole accretion has to fall rapidly to allow galaxies to ``overtake'' their black holes and produce a large change of in the $m_{\rm bh}/m_{*}$ ratio. Such a rapid drop in BH accretion rates is, however, inconsistent with observations (see above). 

In order to overcome this apparent problem, it is clear that one of the underlying assumptions of the simple convolution model must be changed.  As discussed above, the problem arises because all galaxies at high redshift were assumed to contain massive black holes described by the ``high'' $m_{\rm bh}/m_{*}$ ratio. These black holes all built up over essentially the full span of cosmic time.  The obvious solution is thus to have the growth of black holes occurring over a much shorter time interval and only in a fraction of the galaxy population. Crucially, in order to overcome this "memory problem" we need the black hole masses in AGN at any epoch to not be connected with the BH masses in AGN at earlier epochs.  This requires a given galaxy to have an active AGN at only one cosmic epoch even if, during that particular short epoch, it flickers on and off. As we expect most of the AGN activity to happen in galaxies around $M^{*}$ we can choose to make these active galaxies be those that are just about to end their star-forming lives and join the quiescent population.  The observed $m_{\rm bh}/m_{*}$ ratio in AGN at a particular epoch then describes the black holes in these galaxies, which will soon be removed from the AGN population, and will not constrain the masses of the black holes that are be present in other galaxies and which will go through their own active stage at a later epoch.  In other words, $m_{\rm bh}/m_{*}$ becomes a measure of the black holes in the active population only and galaxies that become active at later times will be free to have different $m_{\rm bh}/m_{*}$, irrespective of the value seen at earlier times.

The current paper explores this new co-evolution scenario.  As will be seen, a key point is that by explicitly tying the growth of black holes to a short-lived phase associated in some way to the end of the star-forming life of a galaxy, we can use the \cite{Pen10} formalism to link the number of ``quenching'' galaxies, i.e., the number of AGN, to the cosmic evolution of the star-formation rate of the Universe, thereby bringing the build-up of the stellar populations of galaxies into the model and thereby ``closing the loop'' mentioned earlier.  \par 

\section{Analysis of the new quenching scenario} \label{sec:NewmodelQ}

\begin{figure*}[ht!]
    \centering
    \includegraphics[width=.99\textwidth]{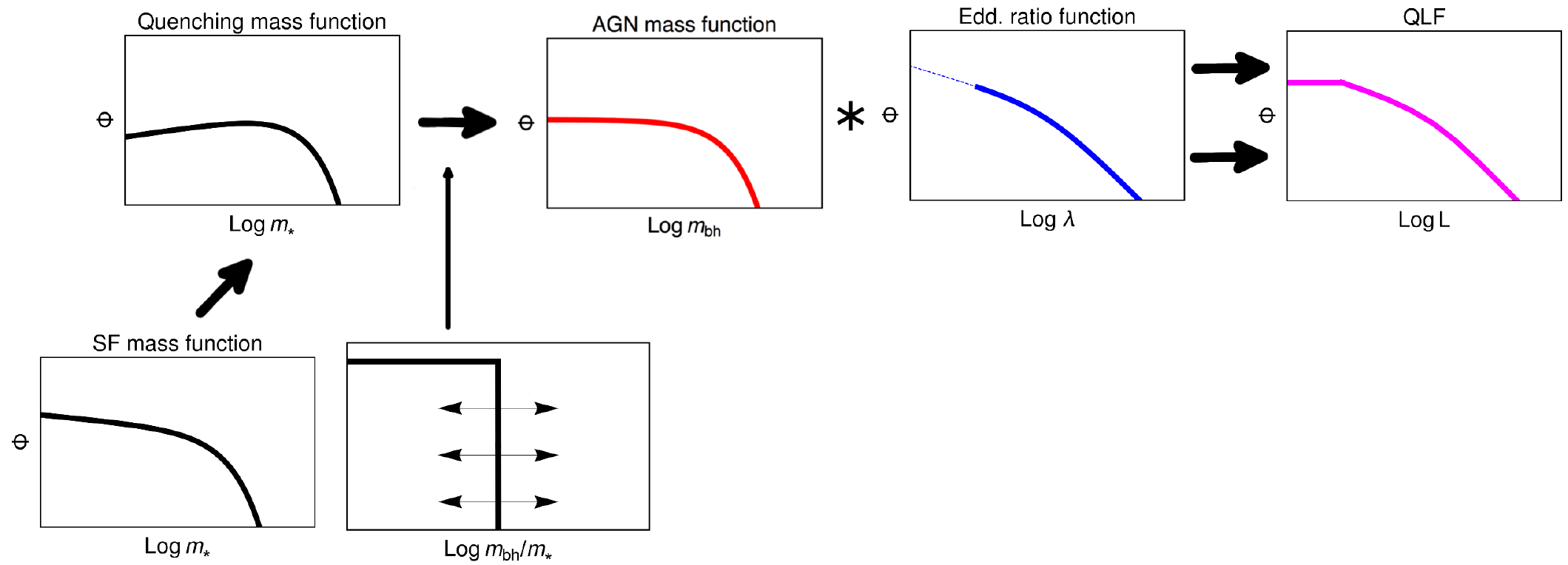}    \caption{Schematic representation of the co-evolution scenario used in this work.  In this scenario only star-forming galaxies that are undergoing (or about to undergo, or just finished) mass-quenching host an AGN. The mass function of ``quenching'' galaxies is derived from the mass function of star-forming galaxies using the formalism from \cite{Pen10}. The AGN mass function is then derived by convolving this ``quenching''  mass function with a step-like function describing the $m_{\rm bh}/m_{*}$ relation, which is flat up to the $m_{\rm bh}/m_{*}$ value at which the galaxies quench, i.e. cease forming stars or accreting on to their black holes. The quasar luminosity function is then given as before by a convolution of the AGN mass function with the Eddington distribution. We also indicate that the Eddington ratio function cannot extended to arbitrarily low values, otherwise the black holes will grow too slowly for the scenario to work. This creates another break at low luminosities, below the ``knee" or characteristic luminosity of the QLF. Compare with Figure \ref{fig:OldModel}. }
    \label{fig:NewModel}
\end{figure*} 

We will now consider a model in which AGN are not continuously growing during the whole time that the host galaxy is building up its stellar population, but in which the entire black hole accretion happens in single, short-lived (ideally almost $\delta$-like), burst of mass accretion just as the galaxy is about to quench its star-formation \cite[as also supported by recent FIRE simulations; see][]{Ang17}. 
We stress that, while we operatively confine the episode of BH growth (and AGN activity) to the period during which the host galaxy experiences quenching, our model does not explicitly assume, nor require, any physical link between the two phenomena, and specifically it does not necessitate that SF quenching is causally due to the AGN activity. 
The crucial component of the model is that AGN host galaxies are removed from the star-forming galaxy population at more or less the same time as their black holes grow in a short-lived AGN phase.
\par  As elaborated in Section \ref{sec:Reprise}, this clearly overcomes the problem of slow co-evolution described above. The black holes now grow over a very short period of time, and the $m_{\rm bh}/m_{*}$ in these objects is unaffected by accretion in other systems that happened at earlier times.  The \textit{current mass ratio} between the total stellar mass and black hole mass is now not given by the integral over the long-term relation between the mass accretion histories of galaxies and corresponding black holes, but with a \textit{current mass accretion rate}. We will explore and further clarify these statements below.

\subsection{Quenching scenario} \label{sec:Newmodel}

We start, as in \citetalias{Cap15}, with considering the mass function of star-forming galaxies (see the schematic illustration in Figure \ref{fig:NewModel}). In \citetalias{Cap15} we assumed  that all star-forming galaxies have equal chance of hosting an AGN.  We now identify, as explained above, the population of star-forming galaxies that are being mass-quenched \citep{Pen10} as the parent population of AGN at a given epoch. \par
We note that \cite{Bon16} already evaluated the consistency of the quenching mass function from \cite{Pen10} with the AGN host galaxy population in the simplified manner. They have used a maximum likelihood approach to jointly fit the stellar mass function and the Eddington ratio distribution from the X-ray luminosity functions. They have showed that the number density and stellar mass distribution at the high mass end of their inferred galaxy population is consistent with the prediction for the quenching population from \citep{Pen10}. In this work we will perform the full evaluation of this statement using the quenching mass function from \citep{Pen10} as the base for the AGN host population. \par 
  
The mass function of ``quenching'' galaxies can be described with the Schechter formula:
\begin{equation} \label{eq:SchQ}
\phi_{Qing}(m_{*})\equiv \frac{dN }{d \log m_{*}} = \phi^{*}_{Qing} \left( \frac{m_{*}}{M^{*}}\right)^{\alpha_{Qing}}  \exp \left(-  \frac{m_{*}}{M^{*}}\right),  
\end{equation}
where $\phi^{*}_{Qing}$ is the normalization, $M^{*}$ is the Schechter mass and $\alpha_{Qing}$ is the low-mass end slope of the quenching population. The parameters of this function are directly connected with the mass function of star-forming galaxies \citep{Pen10}; while the characteristic Schechter mass is the same for both populations, the low mass slope of galaxies that are being quenched differs from the low mass slope of star-forming galaxies by a factor of unity, i.e., $\alpha_{Qing} = \alpha_{SF}+1$. 
To derive the normalization, we can use the fact that the normalization of the mass function of galaxies that are being quenched is proportional to the normalization of the mass function of star-forming galaxies ($\phi^{*}_{SF}$) and their quenching rate ({rsSFR}) \citep{Pen10}.
The fact that the quenching rate is given by the rsSFR is one of the key insights from \cite{Pen10} that forms the basis of the current work. The main observational driver of this conclusion is the fact that the characteristic Schechter $M^{*}$ of the mass function $\phi(m)$ of the star-forming population has been essentially constant out to at least $z \sim 2.5$ (and likely to $z \sim 4$), despite the substantial increase in stellar mass (by a factor of 10-30) of any galaxy that stays on the star-forming Main Sequence over this prolonged time period \citealp{Bel03,Bel07,Ilb10,Poz10,Ilb13}. 
The overall rate of quenching in the population is therefore related to the rate at which galaxies increase in (log) mass, which is given by the rsSFR of the Main Sequence.
The exponential drop-off of the mass function means that for a given star-forming galaxy, the mass-quenching rate, $\eta_{m}$, which reflects the probability that a given galaxy is quenching over a unit time has to be proportional to the rsSFR and the logarithmic gradient of the mass function, which is $m$ for the case of the Schechter function. When these two relations are combined it follows that $\eta_{m} = (M^{*})^{-1} \cdot SFR$, where $M^{*}$ is a constant of proportionality and we recognize it as the Schechter mass of the galaxy mass function.

\par Once when galaxy has entered the quenching population and is therefore considered, in this model, an AGN host galaxy, time spent in this phase is inversely proportional to the speed of the black hole mass growth, i.e., black hole e-folding rate. From this, it follows that
\begin{equation} \label{eq:phiQing}
\phi^{*}_{Qing} = \phi^{*}_{SF} \cdot rsSFR \cdot \tau, 
\end{equation}
 where $\tau$ is black hole mass e-folding time. 
\par

\begin{deluxetable*}{llrr}  
\tabletypesize{\scriptsize}
\tablecaption{Overview of the model parameters \label{tab:parameters}}
\tablewidth{0pt}
\tablehead{Symbol & Description &  Units  & Value }

\startdata
\cutinhead{ Star-forming mass function, Equation \eqref{eq:SchSF}}
$\phi^{*}_{SF}$ &normalization of the SF-galaxy mass function& dex$^{-1}$Mpc$^{-3}$ & Equation \eqref{eq:phiSFWithRedshift}\\
$M^{*}$ &Schechter mass & $M_{\odot}$ & $10^{10.85}$\\
$\alpha_{SF}$ &low-mass end slope of the SF-galaxy mass function & $-$ & -0.45\\
\cutinhead{ Quenching mass function, Equation \eqref{eq:SchQ}}
$\phi^{*}_{Qing}$ &normalization of the quenching-galaxy mass function&  dex$^{-1}$Mpc$^{-3}$  & Equation \eqref{eq:phiQing}\\
$M^{*}$ &Schechter mass& $M_{\odot}$ & $10^{10.85}$\\
$\alpha_{Qing}$ &low-mass end slope of the quenching-galaxy mass function& $-$ & -0.45+1\\
\cutinhead{ Eddington ratio function, Equation \eqref{eq:Ed}}
$\xi^{*}_{\lambda}$ &normalization of the Eddington ratio distribution&  dex$^{-1}$Mpc$^{-3}$  & $\sim \phi^{*}_{QLF}/\phi^{*}_{Qing}$\\
$\lambda^{*}$ &knee of  the Eddington ratio distribution & $M_{\odot}$ & $0.048 (1+z)^{2.5}\mbox{; } z<2$  \\ 
&&& \mbox{const}\mbox{; } $z>2$\\
$\delta_{1}$ &low-Eddington end slope & $-$ & -0.45\\
$\delta_{2}$ &high-Eddington end slope& $-$ & -2.45\\
$\lambda_{min}$ &lower cutoff of the Eddington ratio distribution& $-$ & $\lambda^{*}/10$\\
\cutinhead{Quasar luminosity function, Equation \eqref{eq:QLF}}
$\phi^{*}_{QLF}$ &normalization of  the QLF&  dex$^{-1}$Mpc$^{-3}$  & \cite{Cap15} fit to data from \cite{Hop07}, \\
&&&  also \cite{Ued14} and \cite{Air15}\\
$L^{*}$ &knee of  the QLF& $\mbox{erg}/\mbox{s}$ & \cite{Cap15} fit to data from \cite{Hop07}, \\
&&&  also \cite{Ued14} and \cite{Air15}\\
$\gamma_{1}$ &faint end slope of the QLF& $-$ & 0.45\\
$\gamma_{2}$ &bright end slope of the QLF& $-$ & 2.45\\
\cutinhead{Global parameters}
$\epsilon$ &efficiency of conversion of mass to luminosity in an AGN& - & 0.04\\
$rSSFR|_{ms}$ &reduced specific star formation rate of the Main Sequence& Gyr$^{-1}$ &  0.07 (1+z)$^{2.5}$ \\
$SFRD$ &star-formation density in the Universe& $M_{\Sun}$ yr$^{-1}$ Mpc$^{-3}$ & from \cite{Ilb13} and \cite{Mad14}\\
$BHLD$ &AGN luminosity density in the Universe& $M_{\Sun}$ yr$^{-1}$ Mpc$^{-3}$ &  from \cite{Hop07},\cite{Ued14} and \cite{Air15}\\
\enddata

\end{deluxetable*}

As galaxies enter the quenching population we assume that black holes start growing from some low ``seed'' mass and accrete until they reach a final mass ratio $m_{\rm bh}/m_{*}$ at which point the AGN phase ceases.  If the Eddington ratio distribution is independent of black hole mass, the black holes grow exponentially which corresponds, in logarithmic space, to a convolution of the AGN host mass function with a step-like function with a lower limit set to reproduce the ``seed" AGN mass, and with the upper limit given by the ``final'' mass ratio (see Figure~\ref{fig:NewModel}).  
Since the galaxy quenches star-formation shortly before or shortly after this phase, neither the black hole mass nor the stellar mass will change thereafter, and the limiting $m_{\rm bh}/m_{*}$ will therefore be the value that the galaxy has for the remainder of time.  The limiting $m_{\rm bh}/m_{*}$ of this distribution function therefore sets up the observed $m_{\rm bh}/m_{*}$ relation. This means that the resulting Schechter function describing the black hole mass function of the AGN population, has Schechter mass of $M^{*}_{AGN}= M^{*} \cdot m_{\rm bh}/m_{*}$ where $m_{\rm bh}/m_{*}$ is the upper, ``threshold", value of the distribution of AGN masses. \par

To fully specify the resulting AGN mass function, we note that the low mass slope of the AGN mass function is given by the smaller of two power-law slopes of convolving functions, $\alpha_{AGN}=\mbox{min}(\alpha_{Qing},0)$ (see \citetalias{Cap15} for details). Therefore the AGN mass function will have low mass end slope $\alpha_{AGN}=0$.    \par

Once the AGN mass function has been fully constructed, the QLF is again simply a convolution of the AGN mass function and the Eddington ratio distribution:
\begin{equation} \label{eq:CreateQLFChpThree}
\phi (L,z) = \int \phi_{bh} (m_{AGN},z) \xi (\lambda,z) d \log\lambda,
\end{equation}
where we use $AGN$ when denoting the AGN mass function to emphasize that the AGN mass function used now is different in shape than the one considered in Equation \eqref{eq:CreateQLF}. 
We emphasize that, apart from the difference in the input shape of the AGN mass function, this convolution is exactly the same as the convolution considered in our previous paper reviewed in Section~\ref{sec:Reprise}.  Given this, all of our general conclusions about the connections between the AGN BH mass function, the Eddington ratio distribution function and the QLF, as well as the implications for the form of the $m_{\rm bh}/m_{*}$ relation for passive galaxies, can be carried across to this new scenario as well. \par

A very important point is that the Eddington ratio distribution also determines the duration of the black hole growth phase, i.e., the timescale $\tau$ and thus the value of $\phi_{Qing}^*$. For example, when the mean Eddington ratio is higher, the BHs grow more rapidly, and consequently spend less time in the active phase. Therefore, the normalization of the AGN mass function and thus the normalization of the AGN luminosity function is lowered. It is the introduction of the Eddington ratio distribution into the calculation of densities via the \cite{Pen10} formalism, which includes also the history of star-formation in galaxies, that effectively eliminates the degeneracy that was inherent in \citetalias{Cap15}.

\subsection{New predictions for the mass function of AGN and of their host galaxies} \label{sec:massFunOfAGN}

Because the QLF is the result of the convolution of the AGN black hole mass function and the Eddington ratio distribution, we can immediately calculate the predicted mass function of AGN black holes, and of AGN host galaxies, as a function of their selection luminosity. This is similar to what was described in \citetalias{Cap15} for the case of ``co-existence'' scenario. We will show that the predictions for these two scenarios differ significantly and we will use this extensively when trying to differentiate between observable predictions of the two scenarios in Section \ref{sec:Dif}. \par

\begin{figure*}[htp!]
    \centering
    \includegraphics[width=\textwidth]{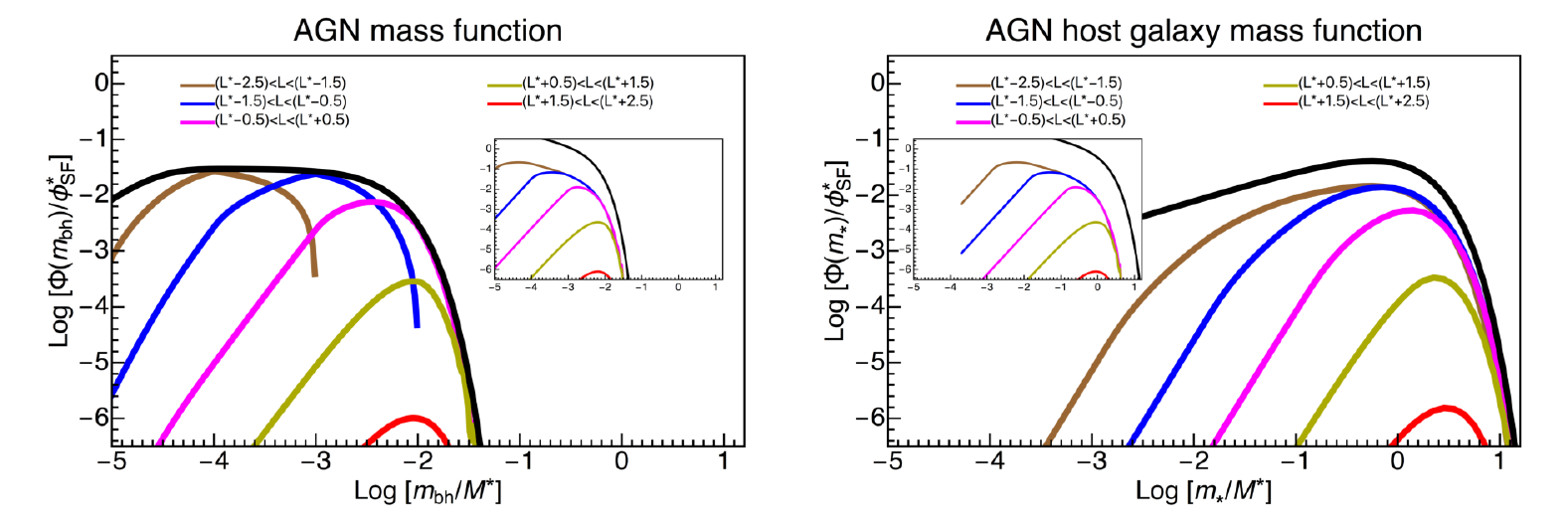}    \caption{Expected shape of the AGN black hole and the AGN host galaxy mass functions, selected in different AGN luminosity bins relative to L*. Masses are plotted relative to the Schechter $M^{*}$ of the galaxy population ($M^{*}\sim 10^{10.85} ~M_{\Sun}$) and $\Phi$ relative to the $\phi^{*}_{SF}$ of the star-forming galaxy population. The insets show the same results that were derived from the scenario presented in \citetalias{Cap15} and are comparable to the Figure 2 in that paper, here presented without scatter for clarity.
$L$ is given in logarithmic units. The black thick lines show the full AGN mass function (left) and AGN host mass function (right). } 
    \label{fig:MFAGNQScenario}
\end{figure*}

To generate the predictions, we have used the AGN mass function described above and an Eddington ratio distribution given by broken power-law function given in Equation \eqref{eq:Ed}, with a lower limit on the Eddington distribution being $\lambda_{min}=\lambda^{*}/10$. We impose this limit to ensure that the AGN accretion mimics a $\delta$-like burst of mass accretion which is a defining feature of the model. The real AGN are unlikely to have such a sharp cutoff at a certain Eddington ratio and this should be considered a simplification which allows the model to be analytically tractable. The exact choice of the lower cutoff is not crucial as long as the process of mass accretion is quick enough.
We do not add, at this point, any additional intrinsic scatter to the galaxy - black hole relation, i.e., to the limiting mass ratio at which the AGN phase is terminated. We normalize all our results to the parameters of the mass function of star-forming galaxies, $\phi^{*}_{SF}$ and $M^{*}$, and also separate results in bins of luminosity relative to the characteristic luminosity of QLF, $L^{*}$. We show our results in Figure \ref{fig:MFAGNQScenario}. \par

We see that for luminosities above $L^{*}$, the effect of the AGN luminosity cut is to change the number of host galaxies, but not to change by much the characteristic mass of the distribution. This is a consequence of the fact that, at these high luminosities, the QLF is dominated by black holes and galaxies that are at the Schechter mass ($M^*$) of their respective mass functions and that are radiating at, or above, the characteristic Eddington ratio ($\lambda^*$). Therefore, the mass function at these luminosities is dominated by an exponential drop off above $M^*$, while below $M^*$ it is characterized with a low mass slope which is given with $\alpha = \alpha_{AGN} + \gamma_{2}$ (see \citetalias{Cap15}). This behavior above $L^*$ is virtually the same as the one presented in \citetalias{Cap15}, as both scenarios use a Schechter mass function as the input AGN mass function. 
The only difference between the two scenarios is therefore in the relatively small change of $\alpha_{AGN}$ slope ($\alpha_{AGN} = \alpha_{SF}  = -0.45$ before and $\alpha_{AGN} \approx 0 $ now). \par

Much more substantial differences arise at luminosities below $L^{*}$. Looking first at the AGN mass functions, in the left panel for Fig.~\ref{fig:MFAGNQScenario} we see that the different AGN luminosities originate from different parts of the AGN mass function. This is a consequence of a fact that our Eddington ratio distribution is quite ``narrow'', so a given AGN luminosity is dominated by black holes in a relatively narrow range of mass. The AGN mass function always reaches its peak at
\begin{equation}
m_{peak,AGN} \sim \frac{L}{10^{38.1}\lambda^{*}},  \hspace{1cm} L<L_{*}.
\end{equation}
We can again recognize that each individual mass function at a given luminosity (bin) has a low mass slope of $\alpha_{AGN}+\gamma_{2}$. This is caused by low mass AGN, described with low mass slope $\alpha_{AGN}$, which are radiating at high Eddington ratios, described with the corresponding high-end slope of the Eddington ratio distribution, $\gamma_{2}$. 
Finally, the AGN mass function drops and disappears at a maximal masses corresponding to of $m_{max,AGN} \simeq L/ (10^{38.1} \lambda_{min})$, reflecting the fact that these black holes are too massive to radiate at these low luminosities, given the assumed $\lambda_{min}$ of the distribution. \par

In case of galaxies at luminosities below $L^{*}$, we can recognize three distinct regimes, similar to the co-existence case presented in \citetalias{Cap15}. At low masses, below $10^{38.1}(m_{\rm bh}/m_{*})^{-1} (\lambda^{*})^{-1}$, the mass function has slope $\alpha_{SF}+1+\gamma_{2}$, and then the slope of $\alpha_{SF}+1$ until the $M^{*}$ mass, after which an exponential drop takes over. \par 

These relations and distributions of galaxies that host AGN of various luminosities are quite different than what we found for the co-existence scenario (presented in \citetalias{Cap15}, compare also with Figure 6. from \cite{Bon16} showing the same information). In this scenario, AGN of all luminosities are mostly likely to be hosted in $M^*$ galaxies. Perhaps surprisingly, even the low luminosity, low mass AGN are most likely to be in massive galaxies, simply because massive galaxies are dominating the quenching population and hosting AGN having a wide range of black hole masses. As a conclusion, in this scenario, while AGN of \textit{different luminosities} are hosted in systems with \textit{different black hole masses}, systems with \textit{similar galaxy mass}, $\sim M^{*}$, dominate the AGN distribution at \textit{all luminosities} because these dominate the quenching population. \par

\begin{figure}[!htp]
    \centering
    \includegraphics[width=.49\textwidth]{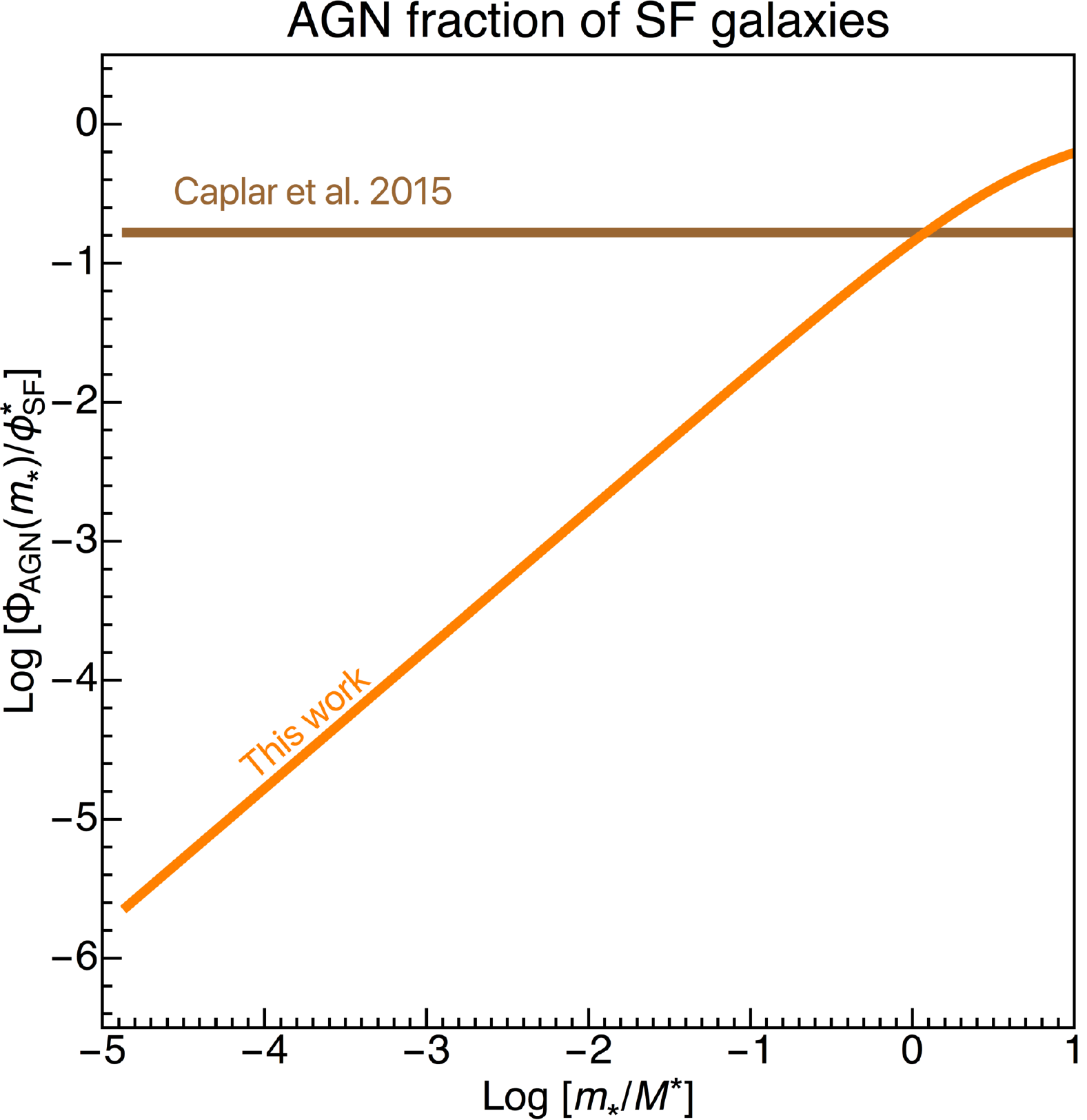}    \caption{ Fraction of star-forming galaxies as a function of stellar mass which are actively accreting at any given time. While in \cite{Cap15} the active fraction was mass independent, in this work the active fraction of star-forming galaxies increases linearly with mass. The normalization of these curves is not so crucial as it depends on the lower cutoff of the Eddington ratio distribution, i.e., what level of accretion in needed to be classified as an AGN.}
    \label{fig:AGNDutyGalaxy}
\end{figure} 

  As a final point in this section, we note the difference in the shape of the mass function of actively accreting galaxies and of their black hole mass function in these two models. The most consequential is  the different shape of the galaxy mass function of the AGN hosts. In \citetalias{Cap15} we postulated that all star-forming galaxies had the same chance to host an AGN, so therefore the mass function of AGN host galaxies had the same shape as the star-forming galaxy mass function, given in Equation \eqref{eq:SchSF}. On the other hand, we have argued that if AGN accretion is connected with quenching we expect the AGN host mass function to be given by the mass function of galaxies undergoing quenching, which is given in Equation \eqref{eq:SchQ}. We show in Figure \ref{fig:AGNDutyGalaxy}, as a function of host galaxy mass, the fraction of star-forming galaxies that we expect to be actively accreting at any given time. We note that although the normalization of these curves is somewhat arbitrary as it depends on the definition of "active" accretion, i.e., on the lower cutoff of the Eddington ratio distribution, there exist significant differences in the mass scaling of the active fraction. We will come back to this point and explore the consequences of this difference in Section \ref{sec:Dif}.

\subsection{The connection between mass accretion and the mass function of galaxies and AGN} \label{sec:con}

The cosmic star formation rate density (SFRD) and the black hole accretion rate density (BHARD) are the basic quantities describing the mass growth of galaxies and black holes in the Universe.  There have been numerous estimates of the SFRD based on observational measurements of the luminosity density of the Universe using different tracers of star-formation. The BHARD can be computed from the observed black hole luminosity density (BHLD) through the radiative efficiency of the gas accretion process, $\epsilon$.
Below we analyze analytically the connections between SFRD and BHLD to find the expected evolution of the active black hole mass function and Eddington ratio distribution. We will use these analytic relations to improve our understanding between different parameters describing SFRD and BHLD, before using measured redshift evolution in Sections \ref{sec:Determination} and \ref{sec:Quant}.   \par 

Both of these integral quantities (SFRD and BHLD) can be derived by knowing the underlying  mass function and the corresponding law connecting mass and luminosity or mass growth. We first derive these relations, i.e., the analytic connections of SFRD and BHARD with the parameters of the appropriate mass functions and mass growth or luminosity relations, before we attempt to connect the redshift evolution of the SFRD and BHLD.  \par

We first consider the SFRD. 
As noted above, the total star formation rate density at any given time in the Universe can be reproduced given the underlying mass function of star-forming galaxies and the law connecting the star-formation rate and the stellar mass of the galaxies. Integrating the Schechter function describing the star-forming mass function over the whole mass range we can derive the total mass of the star-forming population
\begin{equation} \label{eq:rhoSF}
\rho_{SF} = \frac{\phi^{*}_{SF}}{\mbox{ln} 10} \cdot M^{*} \cdot \Gamma(1+\alpha_{SF}),
\end{equation}
where $\Gamma$ is denoting the Gamma function, $\Gamma(x)= \int^{\infty}_{0} t^{x-1} \exp(-t) dt$.  \par

The vast majority of star-formation happens on the Main Sequence of Star Forming Galaxies, a tight linear correlation between the stellar mass and star-formation rate \citep{Bri04,Dad07,Elb07,Noe07}. This means that for galaxies on the Main Sequence, one can define the specific star-formation rate (sSFR), which is not strongly dependent on the stellar mass of a galaxy.  We ignore in this analysis the small tilt in the Main Sequence, i.e., the fact that the correlation between the stellar mass and the star-formation rate is probably not perfectly linear \cite[see][and references therein]{Lil13,Spe14} for simplicity, but also note that this small non-linearity would not make significant changes to our conclusions, since our analysis and results are driven by galaxies in a small range of mass around $M^*$.   \par

The redshift evolution of the star-formation density is then given by the product of the mass density and the characteristic Main Sequence sSFR so it follows that 
\begin{equation} \label{eq:SFRDone}
SFRD  = \frac{\phi^{*}_{SF}}{\mbox{ln} 10} \cdot M^{*} \cdot \Gamma(1+\alpha_{SF})\cdot rsSFR.
\end{equation}
where rsSFR is ``reduced specific star-formation rate" defined in Equation \eqref{eq:rsSFR}.
Given that we know that there is little or no evolution in the $M^{*}$ and the faint end slope $\alpha$ \cite[e.g.,][]{Pen14,Mor15}, the redshift evolution of the SFRD is primarily driven by the evolution of the rsSFR and the parameter  $\phi^{*}_{SF}$ describing the overall normalization of the mass function, i.e.,
\begin{equation} \label{eq:SFRD}
SFRD (z) \propto \phi^{*}_{SF} (z) \cdot  M^{*}\cdot rsSFR(z).
\end{equation}

We now follow the same procedure to investigate the evolution of the BHLD, which can also be derived from the underlying AGN mass function and the mean Eddington ratio. The mass density of the AGN population is given by an equation equivalent to Equation~\eqref{eq:rhoSF}: 
\begin{equation}\begin{split}
\rho_{AGN} &=  \frac{\phi^{*}_{AGN}}{\mbox{ln}10} \cdot M^{*}_{AGN} \cdot \Gamma(1+\alpha_{bh}) \\
& =  \frac{\phi^{*}_{AGN}}{\mbox{ln}10}  \cdot   \left( \frac{m_{\rm bh}}{m_{*}}\right)_{AGN}  M^{*}  \cdot \Gamma(1+\alpha_{bh}),
\end{split}\end{equation}
where in the first step we have denoted the Schechter mass of the AGN mass function with $M^{*}_{AGN}$, and in the second step we have explicitly assumed a linear black hole - galaxy mass relation. \par

The total luminosity of AGN in the Universe is proportional to the total mass of the AGN population and the mean Eddington ratio, $\left\langle \lambda \right\rangle$. Now we can write the equation for the BHLD that is equivalent to Equation~\eqref{eq:SFRDone} for SFRD:
\begin{equation} \label{eq:BHAR1}
BHLD  = 10^{38.1} \cdot\frac{\phi^{*}_{AGN}}{\ln 10}\cdot \left( \frac{m_{\rm bh}}{m_{*}}  \right)_{AGN} M^{*}  \cdot  \Gamma(1+\alpha_{bh}) \cdot \left\langle \lambda \right\rangle .
\end{equation}

We see that the evolution of the BHLD has the same two parameters as the SFRD which influence its redshift dependence, normalization and term describing the mass growth rate (here expressed as Eddington ratio, which is directly connected to the mass growth rate at a given efficiency). The only difference is in the term relating black hole and galaxy mass.  \par

Because the AGN are now only active when the galaxy is mass-quenching, the number of AGN host galaxies is directly proportional to the number of galaxies that are undergoing quenching, $\phi^{*}_{AGN} \propto \phi^{*}_{Qing}$. 
As we have shown in Section \ref{sec:Newmodel}, the normalization of the mass function of galaxies undergoing mass-quenching is given by $\phi^{*}_{Qing} =\phi^{*}_{SF} \cdot rsSFR \cdot \tau $.
Putting this normalization into the equation above leads to
\begin{equation} \label{eq:BHAR1withTau}
BHLD =10^{38.1} \cdot \frac{\phi^{*}_{SF}}{\ln 10} ~ rsSFR ~ \tau \cdot \left( \frac{m_{\rm bh}}{m_{*}}  \right)_{Qing}  M^{*} \cdot\Gamma(1+\alpha_{bh}) \cdot \left\langle \lambda \right\rangle ,
\end{equation}

where we have used the subscript $Qing$ to again remind the reader that the AGN accretion is happening in this scenario only as the galaxies are being quenched. \par

$\tau$ denotes the mean e-folding time for the black holes and we can calculate from the first part of Equation \eqref{eq:rBHAR}  that $\tau = \epsilon/((1-\epsilon)  \cdot (10^{38.1}/10^{37.75})\cdot \langle \lambda \rangle)$. 

Equation~\eqref{eq:BHAR1withTau} can now be simplified by using the expression for $\tau$ to remove the dependence on $\langle \lambda \rangle$. We also use $10^{37.75}$ prefactor to express the $BHLD$ in units of $M_{\Sun}/\mbox{Gyr}$, so that it now reads
\begin{equation} \label{eq:BHAR2}
BHLD = \frac{\phi^{*}_{SF}}{\ln 10}\cdot rsSFR \cdot \left( \frac{m_{\rm bh}}{m_{*}}  \right)_{Qing} M^{*} \cdot \Gamma(1+\alpha_{bh}) \frac{\epsilon}{1-\epsilon}.
\end{equation}
When comparing Equations~\eqref{eq:SFRDone} and \eqref{eq:BHAR2}, we see that the mass ratio in the newly quenched objects (i.e. the limiting mass ratio at which accretion stops) is simply given by the ratio of the BHLD and SFRD multiplied with an efficiency factor, i.e., by the ratio of the mass accretion densities:
\begin{equation} \begin{split}\label{eq:bHARDSFRDTomass}
\left( \frac{m_{\rm bh}}{m_{*}}  \right)_{Qing} &=  \frac{BHLD}{SFRD} \frac{1-\epsilon}{\epsilon}  \\
&\cong \frac{BHARD}{SFRD}.
\end{split}\end{equation}
In the final step above we have ignored the factor $ \Gamma(1+\alpha_{bh}) /  \Gamma(1+\alpha_{SF})$ which is of order of unity. The equation above determines the redshift evolution of the black hole - galaxy mass ratio and therefore the evolution of the characteristic mass of the AGN mass function. In this case, the evolution of the mass ratio is given exactly by the ratio of BHARD and SFRD, i.e., changes in the black hole - galaxy mass ratio reflect, virtually instantaneously, changes in the black hole - galaxy mass accretion ratio and vice versa. \par

Using the same analysis we can also infer the connection of the normalization of the AGN mass function with the normalization of the star-forming galaxy mass function. As we pointed out when deriving Equation~\eqref{eq:BHAR1withTau}, the normalization of the AGN mass function is given by  $\phi^{*}_{AGN} \propto \phi_{SF} \cdot rsSFR \cdot \tau $. Using the fact that $\tau$ and $ \left\langle \lambda \right\rangle$ are inversely related we immediately arrive to
\begin{equation} \label{eq:norm}
\phi^{*}_{AGN} \sim  \phi^{*}_{SF}\cdot \frac{rsSFR}{ \left\langle \lambda \right\rangle}.
\end{equation}
This relation provides a direct link between the normalization of the mass function of star-forming galaxies and that of AGN, which is simply given by the ratio of the mass doubling times.\par
We note that if both the $rsSFR$ and $ \left\langle \lambda \right\rangle$ have the same redshift dependence, then the normalizations of the respective populations will also have the same redshift dependence. This is, in fact, exactly the dependence that is seen at $z \lesssim 2$ \citep{Has05, Cro09, Air10, Air15}. This suggests that $rsSFR$ and $ \left\langle \lambda \right\rangle$ have the same, or similar, redshift dependence over this range of redshifts.
\par

We have now derived the relation for the mass ratio between quenching galaxies and AGN (Equation \ref{eq:bHARDSFRDTomass}) and for the normalization of the AGN mass function (Equation \ref{eq:norm}) which fully defines the evolving AGN mass function. 
To completely describe AGN evolution and its growth we will now derive the evolution of the characteristic Eddington ratio. 
As the $L^{*}$, the characteristic luminosity of the QLF, is proportional to the position of the Schechter-like break in the AGN mass function and the break in the Eddington ratio distribution, the second part of Equation~\eqref{eq:BHAR1} can be recognized as being proportional to $L^{*}$, that is
\begin{equation}
L^{*}\cong10^{38.1} \cdot \left( \frac{m_{\rm bh}}{m_{*}}  \right)_{Qing}  M^{*}  \cdot \lambda^{*} ,
\end{equation}
where we have used the simplifying assumption that $ \left\langle \lambda \right\rangle \propto \lambda^{*}$, i.e., that the characteristic and mean Eddington ratio of the population are linearly related to one another. We can combine the expression above with Equation~\eqref{eq:bHARDSFRDTomass} to express $\lambda^{*}$ fully as a function of observable evolving quantities:
\begin{equation} \label{eq:lambdaEvolution}
\lambda^{*} \cong\frac{SFRD}{BHLD} \cdot \frac{L^{*}}{M^{*}} .
\end{equation}
This expression provides a clear prediction for the evolution of $\lambda^{*}$, which is given by the ratio of two accretion rate densities and by the evolution of the ratio between characteristic QLF luminosity and characteristic galaxy mass. \par

Equations \eqref{eq:bHARDSFRDTomass}, \eqref{eq:norm} and \eqref{eq:lambdaEvolution}  describe the evolution of crucial parameters describing the AGN mass function and the Eddington ratio function, once the evolution of SFRD and BHLD(BHARD) is given. \par

The attractive feature of the model is that both the galaxy mass and black hole mass growth are dominated by objects around $M^{*}$ and $L^{*}$, respectively, allowing us to see these simple scalings between observational quantities purely in terms of the parameters at these crucial points. We stress again the main point that, in this model, the $m_{\rm bh}/m_{*}$ ratio in those systems with AGN reflects, virtually `'`instantaneously'', the cosmic BHLD/SFRD ratio at that particular moment of time. 
This is a consequence of the fact that AGN accrete for a relatively brief time and that all the relevant time-scales are short. 
This is different from the co-existence scenario described in \citetalias{Cap15}, where AGN accreted during the entire time during which the host galaxy is also forming stars. 
This long co-existence means the $m_{\rm bh}/m_{*}$ ratio in the population is set by the integral over the entire accretion history of the host galaxy and AGN, and therefore $m_{\rm bh}/m_{*}$ cannot respond instantaneously to changes in the SFRD/BHARD ratio. 
On the other hand, the relatively short growth time of AGN in the scenario presented here assures that $m_{\rm bh}/m_{*}$ in active systems responds rapidly to changes to mass accretion. \par

\section{Determining the active BH mass and Eddington ratio functions} \label{sec:Determination}

In the previous Section we pointed out the analytical connections between the parameters describing the AGN mass function and the Eddington ratio distribution with the observed QLF and SFRD. Here, we aim to present the main observational constraints on the evolution of these quantities. \par 

For the evolution of the BHARD we use the results presented in \cite{Hop07}, \cite{Ued14}, and \cite{Air15}. 
Even though the first study relies on somewhat dated survey data, it is arguably the most complete as it uses data from the rest-frame optical, soft and hard X-ray, and near- and mid-IR bands, to infer a bolometric AGN luminosity function (i.e., QLF) that is consistent with all the measurements in individual bands, across a wide range in redshift. 
The second and third studies use X-ray surveys with various depths and areas covered to constrain the evolution of BHLD with redshift. X-ray surveys are especially suitable for this purpose since they can also probe (moderately) obscured AGN, and the selection effects are arguably well understood. \par
 
\begin{figure*}[!htb]
    \centering
    \includegraphics[width=.99\textwidth]{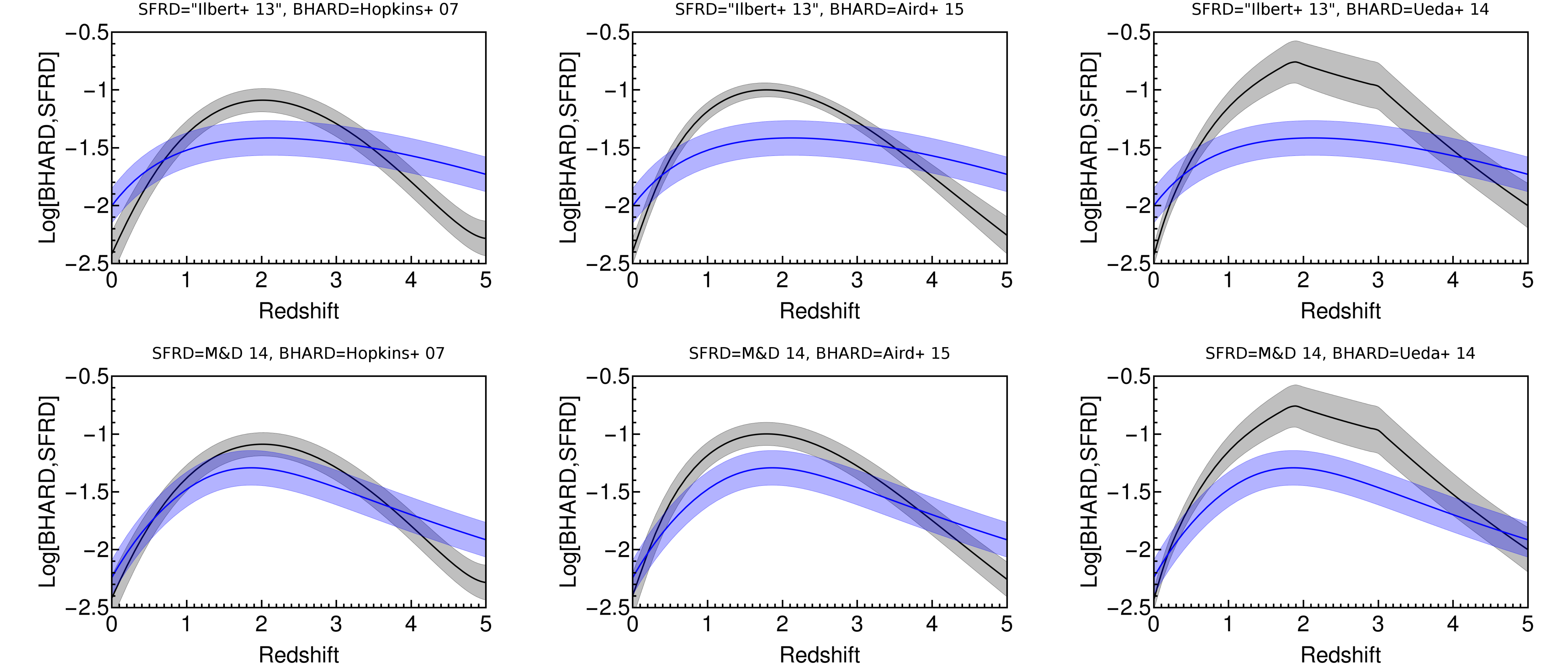}    \caption{Observational determinations of the redshift evolution of the BHARD and SFRD. From left to right, we show the BHARD from \cite{Hop07}, \cite{Air15}, and \cite{Ued14}, while in the top row we show the SFRD derived self-consistently using the model from \cite{Pen10}, assuming that Main Sequence evolves as $rsSFR = 0.07 \cdot (1+z)^{2.5} \mbox{ Gyr}^{-1}$, which reproduces evolution of mass functions in \cite{Ilb13}, while in the bottom row we show the directly determined SFRD from \cite{Mad14}. In each panel the BHARD is shown in black with uncertainties taken from each individual observational study. The BHARD has been derived from the observed BHLD using $\epsilon =0.04$, and has been multiplied by $10^{3}$ to ease the comparison with the much higher SFRD. In each panel the SFRD is shown in blue. The uncertainties are taken to be 0.15 dex and independent of redshift. The units of SFRD and BHARD are $M_{\Sun}$ yr$^{-1}$ Mpc$^{-3}$. }
    \label{fig:BHARDAndSFRD}
\end{figure*} 
 
To derive the BHARD from the observed luminosity densities reported in each of the aforementioned papers, we assume a fixed radiative efficiency of $\epsilon = 0.04$, and adopt the reported observational uncertainties.
We will show in Section~\ref{sec:Quant} that this choice of a low $\epsilon = 0.04$ is most suitable, as it reproduces the mass density of black holes seen in observations. 
Even though some earlier work advocated for higher radiative efficiencies, $\epsilon \simeq 0.1$ \citep{Mer04, Sha09}, the recent re-normalization of the black hole - galaxy relation in the local Universe \citep{Kor13} suggests that lower values of efficiency are needed. 
We will show that our choice of efficiency is consistent with the observations in star-forming and quenched galaxies.\par

We also present results using two different choices for the SFRD. The first one is from \cite{Mad14}, which is based on a compilation of surveys that measured star-formation rates from rest-frame far-ultraviolet or mid- and far-infrared measurements to estimate the evolution of cosmic star-formation density. As noted in \cite{Mad14}, this estimate is slightly inconsistent with measurements of the evolution of the galaxy mass function. 
This could be due to a number of reasons, such as uncertainty in luminosity-dependent dust corrections for UV-measurements, an incorrect initial mass function, the influence of strong nebular lines and other systemic effects. \par

Although \cite{Mad14} is certainly the state-of-the-art compilation of SFRD measurements, as described above, the SFRD enters into our analysis not just to follow the increase of total stellar mass in the Universe but also to have a self-consistent coupling between the Main Sequence rsSFR and the number of mass-quenched objects at each redshift via the \cite{Pen10} formalism. To ensure self-consistency, we therefore also use an SFRD derived using a simplified assumption that the rsSFR of the Main Sequence evolves as $0.07 \, (1+z)^{2.5}\, {\rm Gyr}^{-1}$, and then following through the growth of the galaxy population using the model presented in \cite{Pen10}. This model well reproduces the growth of the galaxy mass function from \cite{Ilb13}, and will enable us later to self-consistently follow the growth of individual galaxies and their black holes, since to derive the number of quenching objects and galaxies that host an ANG, we use the same formalism. 
Since observational estimates of uncertainties are not available for neither of these SFRD determinations, we assume, for plotting purposes only, a representative error of 0.15 dex (at all redshifts).  \par

\begin{figure*}[!htb]
    \centering
    \includegraphics[width=.99\textwidth]{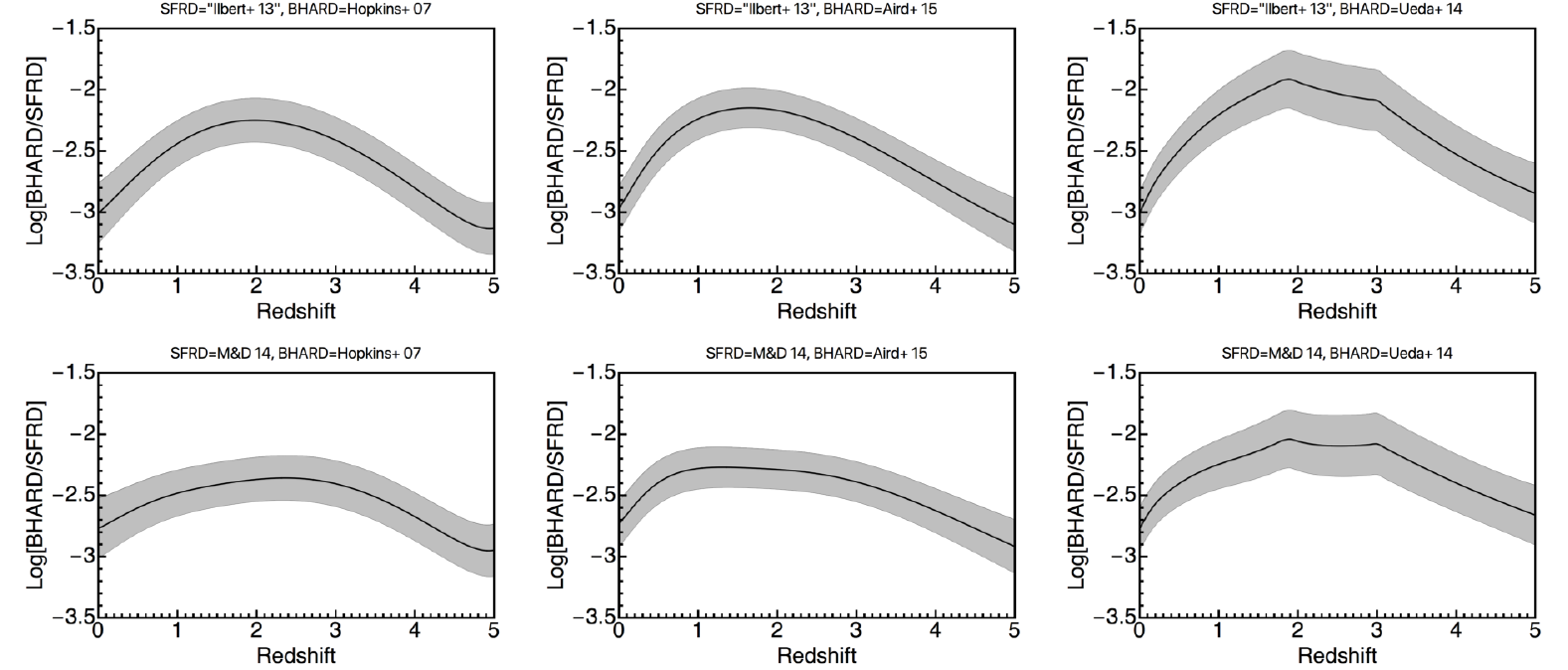}    \caption{Observational determinations of the redshift evolution of the BHARD/SFRD ratio, using the data and panel arrangement shown in Figure 4. This quantity is proportional in the model to $(m_{\rm bh}/m_{*})_{Qing}$, see Equation~\eqref{eq:bHARDSFRDTomass} and text.}
    \label{fig:BHARDSFRD}
\end{figure*}

Figure \ref{fig:BHARDAndSFRD} shows the redshift evolution of the different choices for the BHARD and SFRD, while Figure \ref{fig:BHARDSFRD} shows the redshift evolution of the BHARD/SFRD ratio using the different choices for SFRD and BHARD described above.  
It can be seen that all choices suggest some level of evolution in the BHARD/SFRD ratio.  The ratio is smallest in the local Universe, then rises out to $z \sim 2$ before falling at higher redshifts. The evolution of the BHARD/SFRD ratio points towards an evolution of the limiting $m_{\rm bh}/m_{*}$ ratio in galaxies which are hosting AGN (see Equation~[\ref{eq:bHARDSFRDTomass}]), following roughly $(1+z)^{1.5}$.   
In the scenario presented here, this comes purely from the observed BHARD/SFRD ratio.   In the \citetalias{Cap15} model, this evolutionary trend came from our analysis of the $(m_{\rm bh},L)$ plane of SDSS quasars, which served to break the degeneracy between an evolving $m_{\rm bh}/m_{*}$ and an evolving Eddington ratio distribution in driving the evolution of $L^*$. \par 

We next turn to the evolution of the Eddington ratio distribution. As we argued above (see Equation~\ref{eq:lambdaEvolution}), the characteristic Eddington ratio can be directly deduced (within the framework of this model) from the evolution of the BHARD/SFRD ratio, which we have just established, together with the evolution of the ratio of the characteristic quantities $L^{*}/M^{*}$.
 \par
 
For the evolution of $L^{*}$ in the \cite{Hop07} QLF, we use the results of our ``full'' fit presented in \citetalias{Cap15}.
To estimate the $L^{*}$ evolution in \cite{Air15} we use the results of their own fit to the LADE model. This model describes the QLF evolution with independent evolution of two parameters, corresponding to the typical luminosity and the number density, that is the $L^*$ and $\phi^*$ parameters we use here.
 
The \cite{Air15} work provides best-fitting parameters describing the  evolution of these quantities for the observed hard and soft X-ray data, but unfortunately they do not explicitly provide values for the evolution of the total (combined) X-ray QLF in the LADE model.  
We therefore derive the evolution of $L^{*}$ from the redshift evolution of $\phi_{QLF}$ in the LADE model and the total luminosity density provided in \cite{Air15}. This approach therefore fully account for the evolution of the obscured population. 
We use the $\phi_{QLF}$ evolution in the observed hard band, but we note that the evolution of both normalizations (in both the hard and soft bands) are almost identical and thus this should make little difference in our results. 
The fact that both of these normalizations, which are differently affected by obscuration, have the same redshift evolution makes us confident that the normalization of the total population will behave in the same way. 
Unfortunately, \cite{Ued14} does not parametrize the QLF as a simple broken power law function so we cannot infer the evolution of $L^{*}$ from that study. 
We however note that its results are broadly consistent with the results presented in \cite{Air15}. \par

\begin{figure*}[!htb]
    \centering
    \includegraphics[width=.89\textwidth]{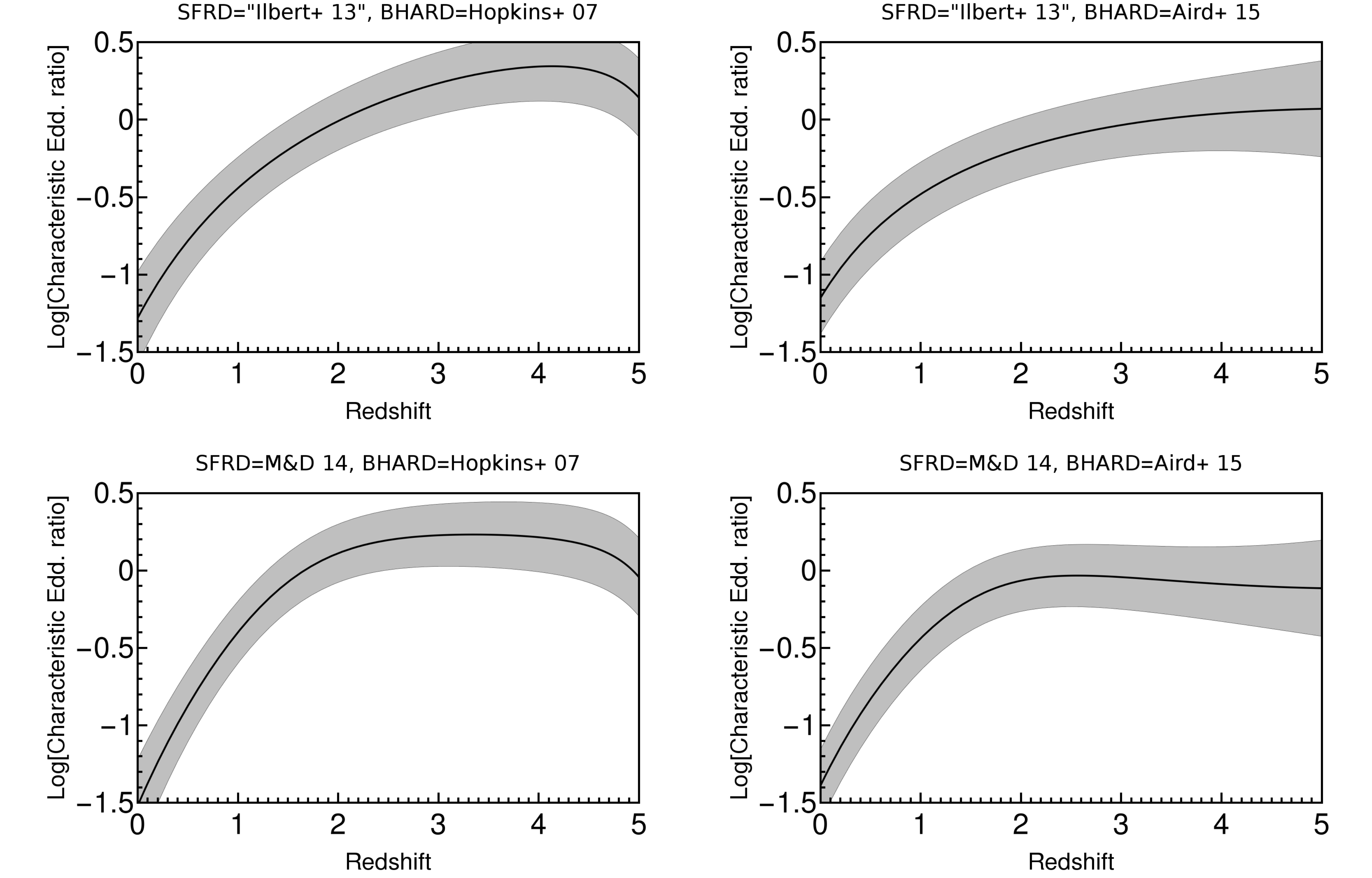}    \caption{The redshift evolution of the characteristic Eddington ratio $\lambda^{*}$ which is determined in the model by $(SFRD/BHARD) \cdot (L^{*}/M^{*})$, see Equation  \eqref{eq:lambdaEvolution} and text. The Figure uses the same panel arrangement as Figure \ref{fig:BHARDSFRD} except that the two rightmost panels are omitted because \cite{Ued14} uses different parametrization of their QLF which does not specify the knee of the QLF, $L^{*}$. 
$\lambda^{*}$ initially rises to $z \sim 2$ as $(1+z)^{2.5}$, interestingly close to the cosmic evolution of the rsSFR of Main Sequence galaxies, after which it flattens off at a value corresponding to the very physically significant Eddington limit, $\lambda^{*} \sim 1$.  Neither of these behaviors was in any way required  by the model. }
    \label{fig:BHARDSFRDLagn}
\end{figure*}

Figure \ref{fig:BHARDSFRDLagn} shows the results for the evolution of the characteristic Eddington ratio $\lambda^*$ that is obtained by combining the evolution of the SFRD/BHARD ratio with the evolution of $L^{*}$ (since $M^*$ is taken to be constant). The typical Eddington ratio rises with redshift, to $z\sim 2$ after which the trend flattens out as $\lambda^*$ saturates at a particular value. The change of the ratio up to $z\sim2$ is around $\sim$1.25 dex, which is what we would expect with the $\lambda^{*} \sim (1+z)^{2.5}$ evolution.   It is immediately apparent that this is very similar to the evolution of the Main Sequence rsSFR relation at these redshifts. Furthermore, we see that this flattening happens as the characteristic Eddington ratio approaches the Eddington limit, i.e., $\lambda^{*} \sim 1$.  This suggests that this is indeed the value above which accretion seems to be rare, and it represents a maximum value at which AGN can shine and accrete mass, at least in the cosmologically relevant sense. 

We stress that the flattening at higher redshifts at this value of $\lambda^{*} \sim 1$ is an output of the model which is driven by the observed form of the changes with redshift of the SFRD, the BHARD, and $L^*$, and is not in any way motivated or dictated by any physical considerations. There were no pre-existing requirements whatsoever in the convolution model which would have either (a) required an evolution of the Eddington ratio with the same dependence as the Main Sequence rsSFR at low redshifts, or (b) produced the saturation at the physically significant Eddington limit at $z \sim 2$.

\section{Quantitative discussion} \label{sec:Quant}

In previous sections, we have set up the model and derived analytic connections between parameters describing the galaxy mass functions, the AGN mass function, the Eddington ratio distribution and the AGN luminosity function (QLF). We have then inferred the redshift evolution of the limiting $m_{\rm bh}/m_{*}$ ratio and of the characteristic Eddington ratio $\lambda^*$, by using observational constraints on the redshift evolution of the BHARD, the SFRD, and $L^{*}$. In this section, we adopt specific choices for the evolutionary trends of all these variables, conduct a quantitative analysis and show our ``predictions'' for the resulting $m_{\rm bh}/m_{*}$ relation in quenched galaxies in the local Universe and for star-forming galaxies at $z < 2$, which can be directly compared with observations.

\subsection{Description of the calculation} \label{DescCalc}

We use the BHARD presented in \cite{Hop07} as we want to compare our results with the evolution of AGN in the BH mass - luminosity plane from SDSS. Because SDSS measurements are the main component of the analysis in \cite{Hop07}, this will ensure that comparisons are done using the same basic dataset. We then also use the ``self-consistent'' SFRD, which uses the rsSFR of the Main Sequence $rsSFR|_{MS}= 0.07 \, (1+z)^{2.5}$ Gyr$^{-1}$ and is prepared as discussed in Section \ref{sec:Determination}. 
As noted, this functional form recreates the growth of the galaxy mass function and the evolution of the number of quenched objects.\par

The model is fully self-consisted because it uses the same inputs to determine both the evolution of the galaxy and AGN populations. As we discussed above, all of the choices for BHARD and SFRD will lead to same qualitative conclusions (see Figures \ref{fig:BHARDSFRD} and \ref{fig:BHARDSFRDLagn}) and the exact choice will not change our main findings. 
\par

Spurred by the results in Section \ref{sec:Determination} and especially those shown in Figure \ref{fig:BHARDSFRDLagn} we adopt a characteristic Eddington ratio which evolves as $\lambda^{*}=0.048 (1+z)^{2.5}$ below redshift 2 and then remains constant at higher redshifts. The fact that $\lambda^{*}$ and $rsSFR$ have the same redshift dependence automatically leads to the same evolution of $\phi^{*}_{SF}$ and $\phi^{*}_{QLF}$ below redshift 2, exactly as observed in the data \citep{Has05, Cro09, Air10, Air15}.

We start our simulation at redshift 6, assuming that all galaxies are star-forming at that redshift and that their mass distribution can be described by a Schechter function. We assume the same non-changing $M^{*}=10^{10.85}\, M_\odot$ and $\alpha_{SF}=-0.45$ seen in the data at lower redshifts. To estimate the normalization at this early epoch, we use the equation (B1) from \cite{Pen12} which provides the relationship for the redshift evolution of the normalization of star-forming galaxies:
\begin{equation} \label{eq:phiSFWithRedshift}
\phi^{*}_{SF}(t)=\phi^{*}_{SF}(t_{0}) e^{\int^{t}_{t_{0}} - (1+ \alpha_{SF})rsSFR(t^{'})dt^{'}}
\end{equation}
and search for redshift 6 normalization which would reproduce the lower-redshift value of $ \phi^{*}_{SF}(z=1)=10^{-2.81}$ Mpc$^{-3}$ dex$^{-1}$ (deduced from the fit presented in \citetalias{Cap15}). We find this to be $\phi^{*}_{SF}(z=6)=10^{-4.58}$ Mpc$^{-3}$ dex$^{-1}$. We then grow galaxies along the Main Sequence in consecutive bins of $\Delta z = 0.002$. In each redshift bin we also deduce which galaxies should quench, according to the mass-quenching law from \cite{Pen10}. \par

There are several ways that one could connect AGN accretion with quenching, either by requiring that AGN accretion happens just before galaxies quench, just after or with some part of accretion happening before and some part happening after quenching, which we assume to be an instantaneous event in the history of an galaxy. In this work, we set up the calculation in such a way that AGN start accreting from their seed mass and reach their final $m_{\rm bh}$, which is given by BHARD/SFRD, at the moment that their host quenches star-formation. This is an operational choice, as both processes (AGN activity and quenching) are unlikely to be well defined, sharp transitions in the life of a galaxy but our results will depend minimally on whether AGN accretion happens just before, during or just after the quenching of star-formation in a galaxy. \par 

When AGN start accreting, we assume that their initial mass function is of the shape as the mass function of ``quenching'' galaxies, i.e., that the low mass slope is $\alpha_{AGN,1} = \alpha_{SF} +1$, where with subscript ``$1$'' we denote that this is the low mass slope of the AGN that have started accreting in a single redshift step of the simulation. 
This choice of slope is the most logical one because is preserves the linear relation between galaxies that are quenching and their black hole mass. This shape is then modified with an assumed log-normal scatter, which we set at $\eta=0.3$ dex for this analysis. To represent the seed mass of black holes, we assume that initial Schechter mass of the AGN mass function is $M^{*}_{AGN,1}=10^6 \, M_{\Sun}$. In each step we then increase the $M^{*}_{AGN,1}$ taking into account the current $\left\langle \lambda \right\rangle$, until the maximal $M^{*}_{AGN,1}$ is reached. As we have shown above, the maximal $M^{*}_{AGN,1}$ is given by the (non-evolving) $M^{*}$ of the galaxy population and the $m_{\rm bh}/m_{*}$ ratio which is reflecting the evolving BHARD/SFRD ratio.   \par

As we discussed in Section \ref{sec:Reprise}, the shape of the QLF is given by the combination of the underlying Eddington ratio distribution and the AGN mass function. The high luminosity end of the QLF is always equivalent to the high-end of the Eddington distribution, while the low luminosity end is given either by the low mass slope of the AGN mass function or the low end slope of the Eddington ratio distribution, whichever is steeper. 
While in \citetalias{Cap15} the low mass slope of the AGN mass function was set to $\alpha_{AGN}=-0.45$ and was therefore steep enough to reproduce the QLF, we have argued in Section \ref{sec:massFunOfAGN} that this is not the case in the present scenario ($\alpha_{AGN} \approx 0$). We thus set the low end slope of the Eddington ratio distribution function to reproduce the low luminosity end of the QLF (i.e., $\delta_{1}=0.45$). 
For simplicity and tractability of the model we set up the high Eddington ratio slope to be constant with redshift. We chose the value representative of the QLF at $z\sim2$ which corresponds to the peak of black hole accretion density and set the high Eddington slope to be $\delta_{2}=-2.45$ \citep{Hop07}. From this, it follows that the QLF will also have non-changing slopes and that the BHARD will only depend on the evolution of the normalization and of the characteristic luminosity. 
We note that considering possible changes of the slopes would minimally affect our results. 
As an example, for a QLF with a given $\phi^{*}_{QLF}$, $L^{*}$ and $\gamma_{1}=0.45$,  varying the high luminosity slope from the lowest value reported in \cite{Hop07}, which is $\gamma_{2}=1.8$ to the highest value, $\gamma_{2}=2.5$, changes the total luminosity density by about 0.09 dex.\par

As we briefly mentioned above, to ensure that AGN growth happens in a quick, approximately $\delta$-like manner, we limit the low end of the Eddington ratio distribution to a fixed fraction of the characteristic Eddington ratio, $\lambda^{*}$.  We set this at a tenth of $\lambda^{*}$, i.e., $\lambda_{min}=\lambda^{*}/10$. 
The choice of terminating the Eddington ratio distribution at a given fraction of $\lambda^{*}$ is the simplest possible, and is attractive as it leads to proportionally between the mean Eddington ratio of the population, $\left\langle  \lambda \right\rangle $, which sets the overall growth timescale, and the characteristic Eddington ratio, $\lambda^{*}$. 
We can then show that the mean Eddington ratio is given by: 
\begin{equation}
\left\langle  \lambda \right\rangle = \frac{\delta_{1}\delta_{2}\left((1+\delta_{2})\lambda_{min} (\frac{\lambda_{min}}{\lambda^{*}})^{\delta_{1}}+(\delta_{1}-\delta_{2}) \right)   }{(1+\delta_{1})(1+\delta_{2}) \left(\delta_{1}+\delta_{2}(-1+(\frac{\lambda_{min}}{\lambda^{*}})^{\delta_{1}}) \right). }
\end{equation}
With this particular choice of parameters, we can use Equations \eqref{eq:rsSFR} and \eqref{eq:rBHAR} and calculate the mass doubling time scale for BH growth. 
We find that it is shorter, by a factor ${\sim}17$, than the mass doubling time scale for galaxies on the Main Sequence. 
This validates our assumption regarding the fast growth of BHs when compared to the typical growth of their star forming host galaxies.\par

\subsection{Predictions of the quenching scenario and comparison with the co-existence scenario } \label{sec:Predictions}

\begin{figure*}[!htb]
    \centering
      \includegraphics[width=.89\textwidth]{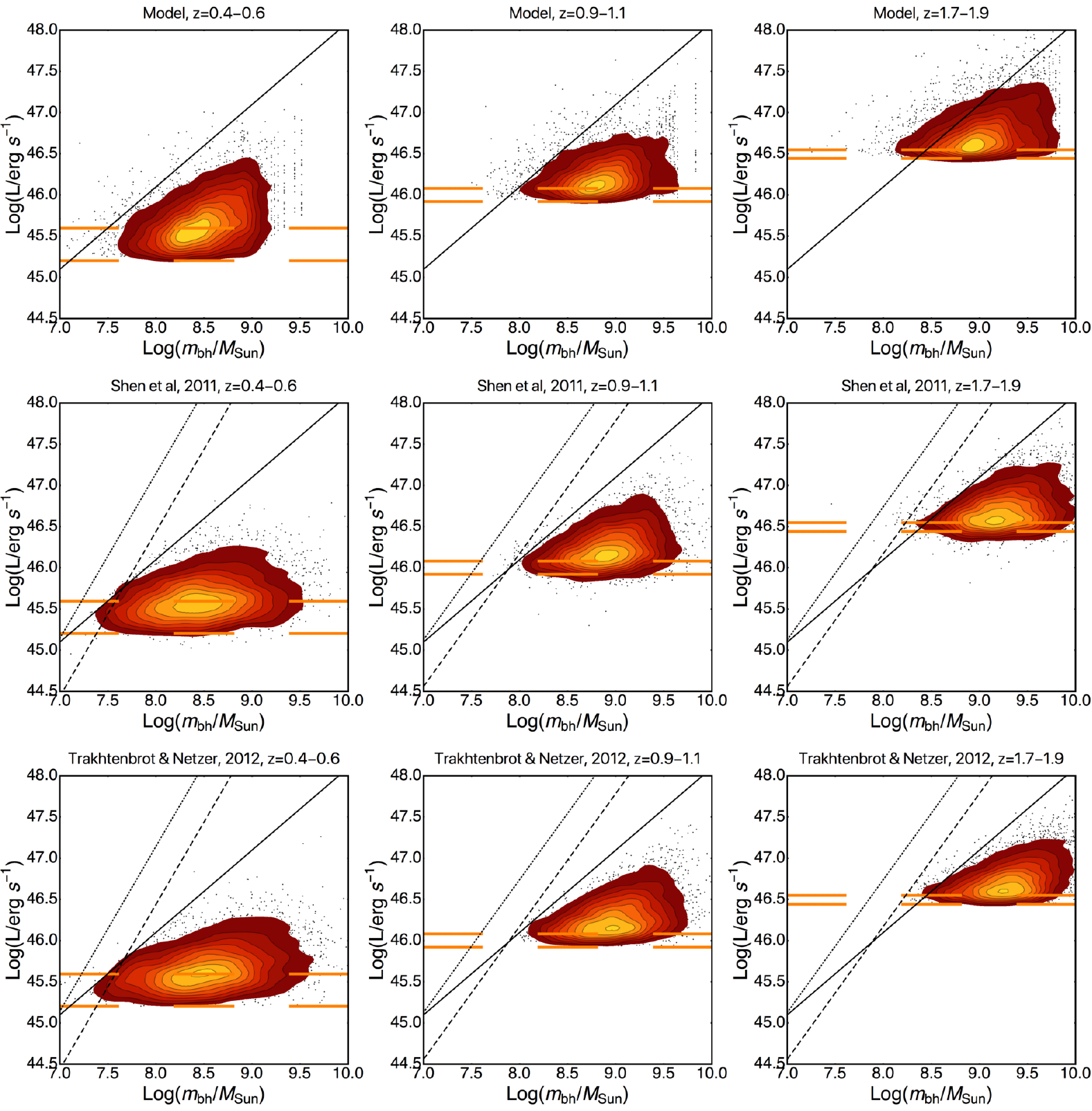} \caption{Top row: the predicted mass-luminosity plane of SDSS quasars in the quenching model, shown in 3 representative redshift bins, left to right. The lower two rows show observational data from \cite{She11} and \cite{Tra12}. In each panel, the thick black line is set at the Eddington limit ($\lambda = 1$), and the dashed orange lines show the calculated luminosity selection limit for the lowest and highest redshift in each bin. The dotted and dashed black lines represent FWHM=1000 km/s and 1500 km/s, respectively, and are shown here to indicate which objects could be missed in observations because of FWHM limit when selecting AGN with broad emission lines. The contours are set at 10$\%$, 20$\%$ etc. values of estimated probability distribution of objects, while the outermost objects are represented as individual dots. The model provides a very good representation o fthe observational data. 
This figure is very similar to Figure 8 in \citetalias{Cap15}. However, whereas in that earlier work, the agreement with SDSS data was achieved by inputing a particular evolution in $m_{\rm bh}/m_{star}$, in the current work this ratio is an output driven by quite different observational data, and this version of the figure is therefore a pure prediction of the quenching model. } 
    \label{fig:MassLum.png}
\end{figure*}

We first present the black hole mass-luminosity plane, generated with the new, quenching scenario we presented in this study, and tuned for comparison with SDSS quasar demographics. We follow exactly the procedure that we explained in detail in \citetalias{Cap15} and which we briefly review here. At each redshift bin we create a mock sample of AGN drawn from the AGN mass function. The number of objects drawn is adjusted to the observed SDSS volume at each redshift bin. We then assign Eddington ratios randomly to each black hole from the (evolving) Eddington ratio distribution function. 
We also apply the obscuration prescription from \cite{Hop07}, to leave only (mock) sources that would be observed by an SDSS-like optical survey. \par

Figure \ref{fig:MassLum.png} shows the evolution of the mass-luminosity plane at three representative redshift bins.
We see that the predicted distributions (top panels) generally follow the observed data (center and bottom panels; taken from \citealt{She11} and \citealt{Tra12}).
The interested reader can compare this with results from \citetalias{Cap15} (Fig.~8 there) and see that the results of the two models are comparable in this sense. 
This is not surprising, and is a consequence of the (required) evolution, in the current model, in the mass ratio of roughly $m_{\rm bh}/m_{*} \propto (1+z)^{1.5}$ and in the Eddington ratio of $\lambda^{*} \propto (1+z)^{2.5}$. These are quite similar to the $(1+z)^{2}$ trends in both quantities which was \emph{postulated} in \citetalias{Cap15} precisely to match the observed $(m_{\rm bh}, L)$ data.  \par 
We stress again that, while in \citetalias{Cap15} we have introduced the evolution of $m_{\rm bh}/m_{*}$ as an input to the model in order to break the degeneracy between the two, \textit{the evolutionary trends in both these quantities are now a direct consequence of the observed galaxy and black hole growth, encapsulated in the SFRD and BHARD functions, respectively}.\par

We then consider the predictions of the model for the black hole - galaxy relation for quenched objects in the local Universe. We compare our prediction with the data from \cite{Kor13} and follow a similar procedure as in \citetalias{Cap15}. The only difference is, while before we used the results from \cite{Bir14} to determine the quenching redshifts of passive galaxies, this information is now incorporated in a self-consistent manner in the model.  Given how well both choices reproduce the overall evolution of the galaxy population, this difference is inconsequential. \par
\begin{figure*}[ht]
    \centering
    \includegraphics[width=.99\textwidth]{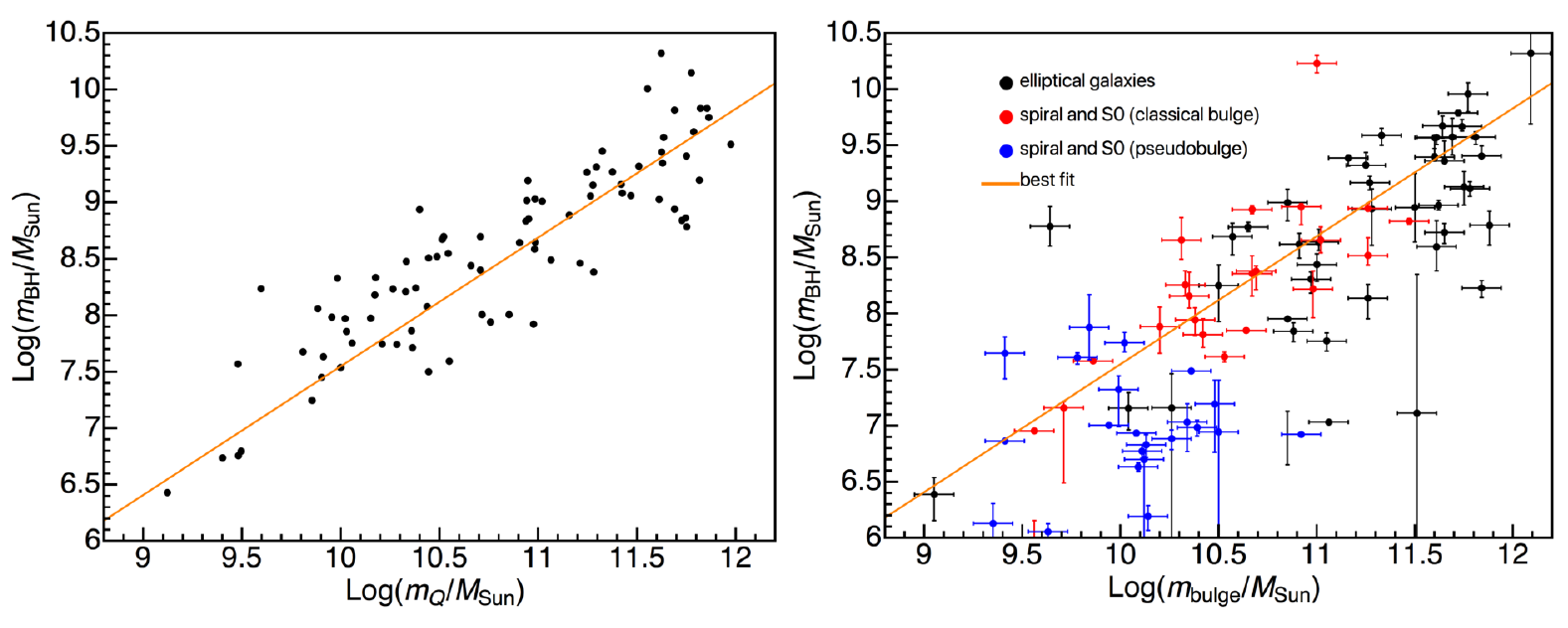}
    \caption{\textit{Left:} The predicted $m_{\rm bh}/m_{*}$ relation in quenched galaxy systems in the local Universe, with artifical data generated by the numerical model. \textit{Right}: The actual $m_{\rm bh}/m_{bulge}$ from \cite{Kor13}. The orange line on both panels shows the best fit to the data reported in \cite{Kor13}. As with our earlier model in \citetalias{Cap15}, the new quenching scenario naturally reproduces the local relation in passive systems.}
    \label{fig:KormQ}
\end{figure*}

In the left panel of Figure \ref{fig:KormQ} we show the expected $m_{\rm bh}/m_{*}$ relation as measured in quenched galaxies in the local Universe, while in the right panel we show the relevant observational data, taken from \cite{Kor13}. We note the excellent agreement between the model predictions and the observations, again similar to what was shown in \citetalias{Cap15}. Given that both models employed the similar information about galaxy evolution and have similar $m_{\rm bh}/m_{*}$ evolution, this result is not surprising. \par

Finally, we also point out that most of the other conclusions that we reached in \citetalias{Cap15}, such as the fact that the ``sub-Eddington" boundary is a simple consequence of the way the AGN population is plotted in this plane, are still also completely valid. That this particular effect is so similar in both cases is simply because the predictions for the two presented scenarios are very similar at high AGN luminosities. SDSS probes luminosities of $L^{*}$ and above and, as we argued when presenting the black hole mass functions at different luminosities in Figure \ref{fig:MFAGNQScenario}, the mass function of black holes at luminosities higher that $L^{*}$ has basically the same shape in the two scenarios that we have considered. Other conclusions from \citetalias{Cap15}, such as the expectation that the $m_{\rm bh}-\sigma$ relation will be redshift independent and the explanation for weaker evolution of the mass ratio at $z \sim 1-2$ in direct observations are similarly not changed. These are both consequence of the mass ratio evolution, which is virtually unchanged in the new scenario, and galaxy evolution, which is unchanged. We elaborate and show this explicitly in Appendix~\ref{sec:bulge}. Finally, we note that since the new scenario was set to also reproduce both the (redshift-resolved) QLF and the BHARD, it also automatically reproduces AGN ``downsizing", even though the Eddington ratio distribution is again independent of the black hole mass, as in \citetalias{Cap15}. \par

\section{Differentiating between the two AGN-galaxy co-evolution scenarios} \label{sec:Dif}

In \citetalias{Cap15} and in this paper, we have presented two models for AGN-galaxy co-evolution based on two different physical scenarios. 
In the first scenario the AGN need not directly influence the galaxy and the co-evolutionary build up of black hole mass and of the stellar populations takes place, in parallel, over the cosmological time-scales. 
In the second scenario AGN are active only around the time at the end of the star-forming life of a galaxy, i.e. either just before, during or just after the (mass-) quenching of star-formation in the galaxy, implying a close physical connection between the two. 
We have seen that both scenarios can reproduce the same overall evolution of the QLF and also explain other features of the observed black hole population in the Universe (including the local relic population). \par

Given that many of the outputs are so similar, reflecting the similar outcomes from the convolutions that are at the heart of the the two models, we might ask whether there are simple observational ways to discriminate between the two models. 
In this Section, we directly compare several predictions of the two scenarios, to construct a diagnostic that can conclusively differentiate between them. We note that the main difference between the scenarios is in which galaxies are assumed to host an AGN. 
In the first model, all star-forming galaxies have an equal chance of hosting an actively accreting black hole and therefore the AGN host mass function has the same shape as the mass function of star-forming galaxies. 
In the second scenario, however, AGN are hosted only in those galaxies that are undergoing (mass-) quenching, and therefore, the mass function of the AGN host galaxies has the same shape as the mass function of quenching galaxies, which in turn has the same shape as that of the population of previously quenched galaxies. 
These two functions differ in their low mass slope by $\Delta\alpha = 1$, i.e., their mass function can be written as
\begin{equation} \label{eq:host}
\phi_{host} (m_*) \propto \left( \frac{m_*}{M^{*}}\right)^{\alpha_{host}}  
\exp \left( - \frac{m_*}{M^{*}} \right) 
 \end{equation}
with $\alpha_{host}$ is $\alpha_{host}=\alpha_{SF}$ in the co-existence scenario and  $\alpha_{host}=\alpha_{SF} +1$  in the quenching scenario (see Equations \ref{eq:SchSF} and \ref{eq:SchQ}). Simply stated, while in the first scenario AGN accretion is equally likely in all star-forming galaxies, in the later scenario AGN activity happens preferentially in more massive star-forming galaxies. \par

In order to differentiate between the models, we thus must consider consider observational studies that simultaneously compare both AGN and galaxy properties.
Such studies are only possible in extragalactic survey fields with rich multi-wavelength data. X-ray surveys offer the best way to trace AGN accretion in an unbiased way and, especially for studies in which we want to investigate AGN-galaxy interdependence, the X-ray coverage is crucial because it is the only band that allows to cleanly identify AGN accretion, as star formation cannot contribute significantly to the sources with luminosity of $L_{x} \gtrsim 10^{42}\,\ergs$ \citep[e.g.,][]{Air17A}. On the other hand mid- and far-infrared photometry are crucial for the characterization of the galaxies' SEDs and particularly their SFRs. The need for this multi-wavelength data restricts such studies to fields with deep X-ray coverage such as \textit{Chandra} Deep Field North (CDF-N; \citealt{Ale03}), \textit{Chandra} Deep Field South (CDF-S; \citealt{Xue11}), COSMOS (XMM-COSMOS, \citealt{Cap09}, C-COSMOS, \citet{Elv09} and the more recently completed \textit{Chandra} COSMOS Legacy Survey, \citealt{Civ16}). 
The mid and far-infrared measurements for these fields with \textit{Spitzer} and \textit{Herschel} are described in \cite{Elb11}, \cite{Lut11} and \cite{Oli12}.

\subsection{Galaxy properties as a function of AGN luminosity}
\label{sec:GalAGNL}

The first quantity we consider is the dependence of the galaxy star-formation rate as a function of the AGN luminosity. For this experiment, we generate another AGN sample using the same method as it has been described in Section \ref{sec:Predictions}, but considering the full luminosity range (without SDSS luminosity cut, as in Section \ref{sec:Predictions}). The AGN sample is divided into luminosity bins, and then the mean star-formation of the AGN hosts is determined. Assuming that the AGN hosts are Main Sequence star-forming galaxies that obey a linear relationship between their mass and star-formation, determining the mean star-formation rate is largely equivalent to determining the mean mass of the host galaxies. As we have argued in Section \ref{sec:massFunOfAGN}, this dependence is intrinsically different in the two AGN-galaxy co-evolution scenarios: while in the co-existence scenario the typical galaxy mass of AGN hosts was dependent on the AGN luminosity, in the quenching scenario galaxies near $\sim M^{*}$ dominate the host mass function at all AGN luminosities.  \par

\begin{figure*}[!htp]
    \centering
    \includegraphics[width=.99\textwidth]{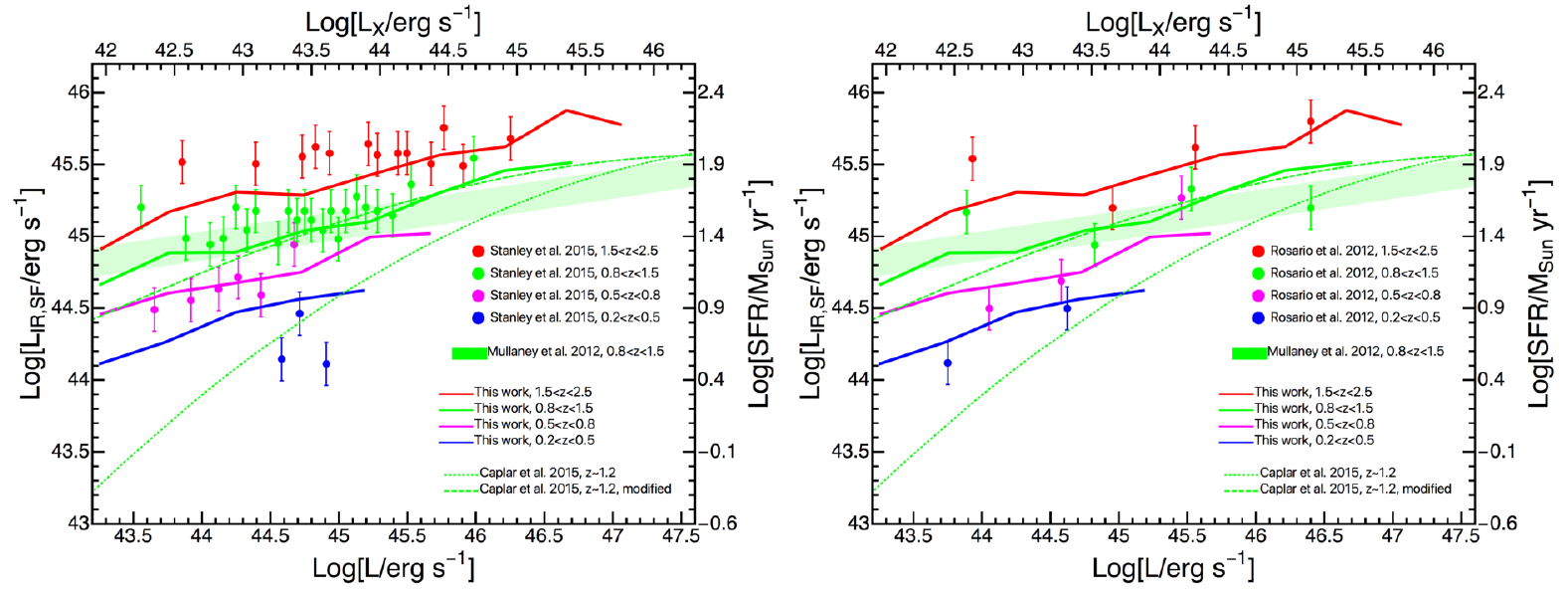}
    \caption{The mean star formation rate in AGN hosts as a function of AGN luminosity. In both panels, the full lines show predictions of our model at four different redshifts, the light green filled area shows the observational results at $0.8 < z < 1.5$ from \citet[][]{Mul12} and the dotted line shows the prediction from the co-existence scenario as presented in \citetalias{Cap15}. 
We also show, with a dashed line, the prediction from a modified version of the \citetalias{Cap15} model, but with a low Eddington ratio slope set to $\delta_{1}=-0.45$. 
In the left panel we also add comparison with the color-matched data points from \citet[][]{Sta15}, while in the right hand panel we show the color-matched data points from \citet[][]{Ros12}. We show some of the results only at redshift $z \simeq 1.2$ to improve clarity of the plot.}   
    \label{fig:SFL2L}
\end{figure*}

We show the results in Figure~\ref{fig:SFL2L}. The data points show observational results from \cite{Ros12} and from \cite{Sta15}. These are directly comparable, but we avoided putting them on the same panel to improve the clarity of the presentation. 
We also show results from \cite{Mul12MNRAS}, using their Equation (4), which connects stellar mass and the AGN $L_{X}$. We transform stellar mass to SFR assuming the Main Sequence from Equation \eqref{eq:lillySFR} and with scatter of 0.3 dex. We show the result as a band with a width of 0.2 dex, corresponding to the scatter seen in the data. 
In both panels, the dual x-axes mark both X-ray and bolometric luminosities. 
We follow the suggestion from \cite{Mul12} and use, for the sake of simplicity, a constant bolometric correction of $k=22.4$ \citep{Vas07}. 
The dual y-axes mark both the IR luminosity due to star-formation, and the equivalent SFR, using the constant scaling in which SFR of 1 $M_{\sun}/{\rm yr}$ corresponds to $10^{43.6}\,{\rm erg s}^{-1}$, as used in \cite{Sta15}. 
The solid lines trace the predictions using the model developed in this paper, in which AGN are connected with quenching, in four different redshifts bins. 
These can be compared to the prediction of the scenario presented in \citetalias{Cap15}, shown with dotted line.  
We also show a modified version of the \citetalias{Cap15} model, in which we set the low Eddington ratio slope to $\delta_{1}=-0.45$, instead of $\delta_{1}=0$ as used in the original work.
This modified model is meant to resemble some of the scenarios discussed in \cite{Sta15}(see more discussion below).

For the sake of clarity, we show only those outcomes from \citetalias{Cap15} and \citealt{Mul12MNRAS} that correspond to redshift $z\simeq 1.2$, and note that the conclusions we draw are applicable to any redshift. \par

As Figure \ref{fig:SFL2L} shows, the quenching model provides an excellent agreement with observations. The predicted slope is almost flat as a consequence of the fact that, in any AGN luminosity bin, the sample is dominated by galaxies at the same stellar mass, i.e., around the characteristic Schechter $M^*$, as seen on Figure \ref{fig:MFAGNQScenario}). 
Even though different luminosity bins have very different black hole masses, these are all hosted by galaxies that are undergoing quenching, which are overwhelmingly massive galaxies around $M^{*}$. 
There is a slight tilt to the predicted relations in both panels. 
This is a consequence of the fact that black hole accretion is not instantaneous - while the black holes are growing through accretion, their galaxies are also growing through star-formation, albeit at a slower pace. This induces a weak dependence in which more luminous objects, corresponding to the more massive black holes, do actually reside in slightly more massive galaxies. \par 

In contrast, the original co-existence scenario from \citetalias{Cap15} shows a very different, much stronger SFR-$L_{\rm X}$ relation. 
As discussed in Section~\ref{sec:massFunOfAGN}, the dependence of host mass on AGN luminosity is quite different below and above $L^{*}$. Below $L^{*}$, the link between galaxy mass and AGN luminosity is linear, while above $L^{*}$ such a scaling cannot continue given the steep drop in the galaxy mass function, and all AGN are effectively hosted in galaxies with roughly $M^{*}$. This can be seen in Figures \ref{fig:SFL2L} as the dotted line flattens its linear dependence at higher luminosities, approaching what is seen in the quenching scenario (i.e., the present work). This line of reasoning is very similar to the reasoning used by \cite{Ber18} who argued that the flat SFR-$L_{\rm X}$ relation is an indicator for the existence of the mass dependence in the Eddington ratio distribution of AGN. \par

The origin of this strong dependence in \citetalias{Cap15} was the rather narrow range of galaxy masses which host AGN of certain luminosity. We note that we can extend this range by modifying $\delta_{1}$, i.e., the low-end slope of the Eddington distribution function. The only constraint from the QLF data is that this slope has to be shallower than the low-luminosity slope of the QLF, which is roughly $\gamma_{1} \simeq 0.45$ \cite[e.g.,][]{Air15}. 
In \citetalias{Cap15} we have used $\delta_{1} =0$ due to its attractive property of describing an exponentially decaying AGN population, but any $\delta_{1} \gtrsim -0.45$ is capable of explaining the data. 
As noted above, Figure \ref{fig:SFL2L} shows the prediction of from a ``modified'' version of the \citetalias{Cap15} model, this time indeed assuming $\delta_{1}=-0.45$. 
We see that this allows for a much better agreement with the data, compared to the original \citetalias{Cap15} model. 
Specifically, high-mass galaxies and their high-mass black holes again dominate the AGN population, especially at higher luminosities. Deviations at low luminosities start to appear due to the fact that the Eddington ratio distribution has to be truncated at a lower cutoff value, as the integral of the distribution cannot exceed unity. 
We note that the differences between these two models are indeed relatively small within the range of luminosities covered by the data.

\subsection{AGN properties as a function of galaxy mass }

An alternative view on the AGN - galaxy connection can be obtained if the data are not binned in bins of AGN luminosity as in the previous subsection, but rather in bins of galaxy mass.  In observational analyses that follow this approach, all star-forming galaxies of a certain stellar mass are considered and then the mean AGN luminosity is calculated by stacking the total X-ray flux at the positions of these objects, including X-ray flux that would not be considered sufficient to claim a detection of an individual galaxy as an AGN. 
This approach is of interest because, even though the underlying data are the same, the different projections can expose different trends in the population (e.g., \citealt{Vol15}). 
Additionally, when selecting objects above a certain AGN luminosity, as above, we are sensitive only to objects that are currently accreting. As we have pointed out above, the main difference between the two AGN-galaxy scenarios was in the mass dependence of the fraction of galaxies that host an AGN. While the first approach was actually testing the average host mass of AGN of different luminosities, this second analysis approach is directly probing mass dependence of the fraction of galaxies hosting an AGN, as the mean luminosity of AGN in a bin containing all star-forming galaxies of a similar mass is obviously sensitive to the fraction of these galaxies that are hosting an AGN.  \par

We now compare the results of our two models with two sets of observations. The first analysis mimics the results one would expect from a 0.9 deg$^{2}$ C-COSMOS field, analyzed in \cite{Rod15}. To do so, we use the results from our analysis in Section~\ref{sec:Quant} and create very large numbers ($10^5$) of AGN hosted in galaxies with $m_{*}>10^{8}~M_{\Sun}$, together with their black hole and galaxy masses, in redshift bins separated by $\Delta z =$ 0.1. We note that in our model this is  largely equivalent to the selection in SFR for star-forming objects, given the assumed linear tight relation between stellar mass and star-formation.
We estimate the number of objects that we expect in each redshift bin in a 0.9 deg$^{2}$ field and draw objects randomly. For the sample at $z\sim 2$ we consider objects between $1.5<z<2.5$, and for the sample at $z\sim 1$ we take objects between $0.5<z<1.5$. \par

The second analysis mimics results from the smaller 0.11 deg$^{2}$ CDF-S field, analyzed in \cite{Mul12} and \cite{Yan17}. We follow the same procedure as described above. 
For the sample at $z\sim 2$ we simulate objects between $1.3<z<2.0$, and for the sample at $z\sim 1$ we simulate objects between $0.5<z<1.3$, as has been done in \cite{Yan17}. 

\begin{figure*}[t] 
    \centering
    \includegraphics[width=.99\textwidth]{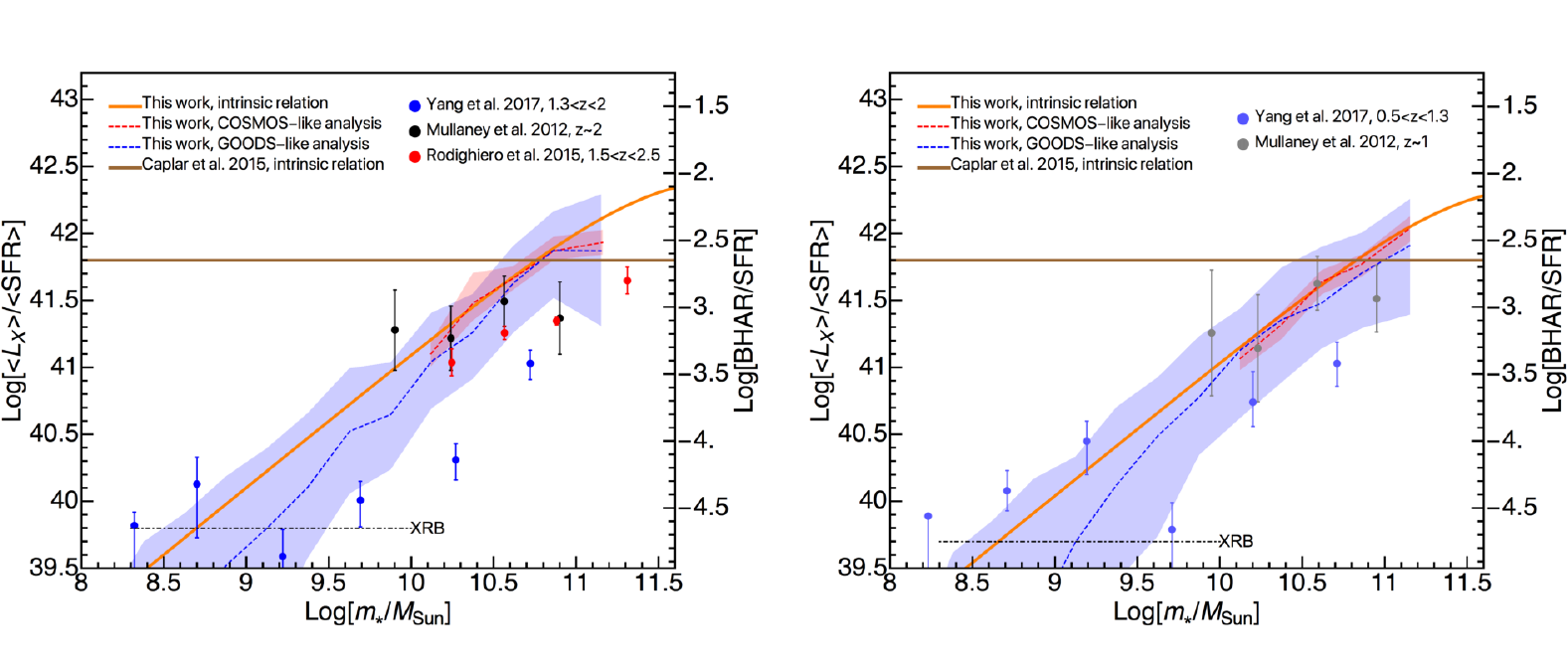}
    \caption{The ratio of the mean X-ray luminosity and mean SFR, $\left\langle  L_{x} \right\rangle / \left\langle SFR \right\rangle$, of star-forming galaxies as a function of their stellar mass. Left panel shows results at $z\sim 1$ and the right panel shows results at $z\sim 2$.  Data is shown from \cite{Yan17}, \cite{Rod15}, and \cite{Mul12}.  The orange line is the prediction from our quenching model assuming an infinitely large survey area. The red and blue lines and shaded areas show the predictions for surveys limited in area, and show COSMOS-like and GOODS-like surveys (see text for more details).  The horizontal brown line is the prediction from our previous model in \citetalias{Cap15}. Compare with Figure \ref{fig:AGNDutyGalaxy}. This Figure prefers the newer model developed in this paper. } 
    \label{fig:LxAsFunOfMass1NoEff}
\end{figure*}

We show the results of these analyses in Figure \ref{fig:LxAsFunOfMass1NoEff}. 
The left panel shows the results at higher redshift of $z\sim 2$, while the right panel shows the results at the lower redshift of $z\sim 1$. 
The different points correspond to the results of observational studies: 
the black and gray points show the CDF-S results by \cite{Mul12}; 
the light blue and blue points show CDF-S results by \cite{Yan17}; 
and the red points show the COSMOS results presented in \cite{Rod15}. 
We also show, as a dashed black line, the level at which the contribution from the unresolved X-ray background becomes appreciable, as estimated by \cite{Yan17}. This serves as a lower limit for estimating the $\langle  L_{x} \rangle / \langle SFR \rangle$ dependence. \par
We have made the X-ray luminosities obtained from these different surveys consistent by following the same prescription as the one described in \cite{Rod15}. The thick brown line shows the dependence with mass that is expected from the co-existence scenario that was presented in \citetalias{Cap15}. The fact that this line is horizontal is a natural consequence of that model because all star-forming galaxies have the same Eddington ratio distribution at all masses and therefore an equal probability of hosting an AGN.  This means that that there is a direct linear correlation between the stellar mass and the mean AGN luminosity measured at that stellar mass. When coupled with the linear correlation between stellar mass and the mean star-formation (i.e., under assumption of linear Main Sequence), this leads to constant relation between the mean stellar mass of the sample and its mean AGN luminosity. \par
On the other hand, the diagonal thick orange line shows the idealized expectation of the mass dependence from the quenching scenario considered in the current work, in which the probability of hosting an AGN increases linearly with galaxy mass, following the increasing likelihood of quenching. The line saturates at high masses as the fraction of star-forming galaxies that are ``quenching'' starts to approach unity. The red and blue lines show our expectations for observed relations in actual surveys. The shaded regions were created by running the simulation of the results from the quenching model 100 times and selecting the 16th and 84th quantiles to indicate the possible spread of the results. Because the COSMOS field is much larger and therefore contains more galaxies, the spread is naturally smaller, as seen in the figures. 
We also show in Appendix \ref{sec:Small}, in qualitative fashion, how the fact that these fields are still relatively small and lack very luminous objects can bias these results and modify the inferred mass dependence. This effect is particularity pronounced when evaluating the results from the small CDF-S field. The exact treatment of this effect is beyond the scope of this paper but we note that it would only strengthen our conclusions. \par

We see from Figure \ref{fig:LxAsFunOfMass1NoEff} that the quenching model presented in this paper provides a much better agreement with the observational data. The difference is particularly pronounced at lower galaxy masses, where the predictions from the two models diverge. We see that the data from \cite{Yan17} suggests consistently lower $\langle L_{x}\rangle / \langle {\rm SFR} \rangle$ compared with measurements of higher-mass galaxies, in accordance with our predictions. \cite{Yan17} notes that the contribution of X-ray binaries (XRBs) to the X-ray signal, not explicitly modeled here, acts to flatten out the dependence at low stellar masses. This explains why the data at lowest stellar masses do not continue to follow the linear trend with stellar mass. \par

We can also comment briefly on the difference in normalization between the different data samples and predictions. The aforementioned observational studies have been done on different datasets and using different methods to estimate SFR.  
Even for data at similar masses, the spread is indicative of systematic differences in estimating  $\langle  L_{X} \rangle / \langle {\rm SFR} \rangle$  ratio. Given these large uncertainties we believe it would be unwise to attach too much significance to the overall normalization of the results, and we emphasize the difference in mass dependence of the $\langle  L_{X} \rangle / \langle SFR \rangle$ relation. \par

In summary, this analysis clearly shows a preference for the mass dependence in the $\langle  L_{X} \rangle / \langle SFR \rangle$ relation, which is consistent with the mass-quenching scenario we present here. Coupled with the results showing links between galaxy properties and AGN luminosity, shown in Section \ref{sec:GalAGNL}, we see that the quenching co-evolution scenario explored in this paper is clearly preferred by the data over the co-existence co-evolution scenario presented in \citetalias{Cap15}. 

\section{Summary} \label{sec:Sum}

We have presented an observationally motivated, phenomenological model which connects the evolving galaxy population with the evolving AGN population. The paper represents an extension of the formalism that we first presented in \citetalias{Cap15}. In the current paper we have explored a scenario in which black holes do not grow over the whole period that the galaxy is forming stars, as in \citetalias{Cap15}, but rather that the central black holes are only active and grow in one relatively short episode as their host galaxies are approaching Schechter mass (M$^{*}$). We operationally connect this process with galaxies undergoing quenching (and more specifically, the so-called ``mass-quenching'') process at the end of their star-forming lives. 
The development of this new model was motivated by the fact that the significant changes, by a factor of ten, in the $m_{\rm bh}/m_{*}$ black hole - galaxy mass ratio between redshifts 0 and 2 that were needed to reproduce observations in the SDSS BH mass- quasar luminosity plane, were barely consistent in the co-existence scenario presented in \citetalias{Cap15} with the observed mass growth rates of the stellar and black hole components. We have then showed analytically the following results:

\begin{enumerate}
\item The black hole - galaxy mass scaling relation for those galaxies that are just about to quench (and shut down further growth of both stellar and black hole mass) is given explicitly by the global BHARD/SFRD ratio of the Universe at the epoch in question. Observational data on the evolution of the BHARD and SFRD suggest that the BHARD/SFRD ratio of these ``quenching galaxies'' was higher at redshift 2 than today by 0.5 to 1 dex, i.e., an evolution of $\sim(1+z)^{1.5}$. This should be the change in the $m_{\rm bh}/m_{*}$ ratio in galaxies that host an AGN. 

\item The evolution of the characteristic Eddington ratio, $\lambda^{*}$, is given by the product of the evolution of the global SFRD/BHARD ratio and the evolution of the characteristic luminosity of the QLF, $L^{*}$ (since it is assumed from observations that the evolution of the Schechter $M^*$ of star-forming galaxies is negligible).   
The evolution of these quantities suggests a rise of $\lambda^{*}$ following $\sim(1+z)^{2.5}$ up to $z\sim 2$. This is strikingly similar to the observed evolution of the cosmic rsSFR.  
At higher redshifts, the observations indicate that the evolution of $\lambda^{*}$ stabilizes at a value close to the physically highly significant Eddington limit, i.e. $\lambda^{*} \sim 1$.

\item The normalization of the AGN mass function, $\phi^{*}_{AGN}$, is proportional to the product of the $\phi^{*}_{SF}$ normalization of the star-forming galaxy mass function and the ratio of the characteristic specific star formation rate of the Main Sequence and the black hole growth rate $\lambda^{*}$. If the rsSFR and $\lambda^{*}$ have the same redshift dependence, as our analysis (in the framework of this model) indicates they do out to $z\sim 2$, the ratio between $\phi^{*}_{AGN}$ and $\phi^{*}_{SF}$ stays constant with time, as observed.

\end{enumerate} 

We have shown that this model, in which AGN activity is connected with galaxy quenching, also reproduces the SDSS BH mass -- quasar luminosity plane, and the observed $m_{\rm bh}/m_{*}$ relation of quenched galaxies in the local Universe. Similarly, other attractive conclusions from \citetalias{Cap15} based on the evolution of the $m_{\rm bh}/m_{*}$ will also hold in the new scenario. These include (i) the explanation for the so-called ``sub-Eddington boundary'', (ii) the reproduction  of downsizing even though the Eddington ratio distribution is independent of mass, (iii) the explanation for the weaker BH-to-host mass ratio evolution at $z\sim 1$ in observations, and (iv) the expectation that the $m_{\rm bh}-\sigma$ relation will be roughly redshift independent.  \par 

This is not surprising as the $m_{\rm bh}/m_{*}$ and $\lambda^{*}$ that we \textit{derived} from observations in the current work are very similar to the evolution that we \textit{postulated} in \citetalias{Cap15}. This congruence was by no means guaranteed, nor set up beforehand, but is a natural consequence, within the framework of this model, of the observed evolution of the BHLD(BHARD), SFRD and the galaxy population.  \par

We have pointed out that the main difference between the co-existence and quenching co-evolution scenarios is in the mass function of galaxies that are hosting AGN. We have therefore conducted a comparison of the two scenarios with observational studies that follow simultaneously the evolution of the mean properties of both galaxy and AGN populations. 
Our main conclusions from this comparison are:

\begin{enumerate}
\item The mean star-formation in AGN host galaxies is only weakly dependent on the luminosity of AGN in the quenching scenario, which is in excellent agreement with the observed relations from \cite{Ros12}, \cite{Mul12MNRAS}, and \cite{Rod15} that show little luminosity dependence in the observed range.  The co-existence scenario as described in \citetalias{Cap15}, predicts a linear correlation between the star-formation rate and AGN luminosity in the luminosity range covered by the observations, but can be made more consistent with observations by modifying low-end Eddington slope dependence. 

\item We make predictions for the mean AGN luminosity/SFR ratio as a function of galaxy mass in the quenching and co-existence scenarios in COSMOS and CDF-S fields. The quenching scenario predicts a linear mass dependence, a consequence of the fact that more massive galaxies are more likely to host an AGN, while the co-existence scenario predicts no mass dependence in the mean AGN luminosity to SFR ratio, since all galaxies are equally likely to host an AGN.  We find that the quenching scenario offers a better description than the co-evolution scenario for the observed galaxy mass dependence of $\langle  L_{X} \rangle / \langle SFR \rangle$ in \cite{Mul12}, \cite{Rod15} and \cite{Yan17}. 
\end{enumerate}  

This work has been supported by the Swiss National Science Foundation.

\bibliography{AGNEvolutionII}

\appendix

\section{The $m_{\rm bh}/m_{bulge}$ correlation in quenched systems }  \label{sec:bulge}

In this section we will further expand our analysis of the scatter in the relations connecting galaxy properties and black holes masses in passive (quenched) galaxies. We have already pointed out that the size evolution of galaxies will act to reduce the scatter in the  
$m_{\rm bh}-\sigma$ relationship, compared to the $m_{\rm bh}/m_{*}$ relation. In this appendix, we now consider the scatter in the $m_{\rm bh}/m_{bulge}$ relation. \par

How the bulges are formed and what dynamical processes are drivers of the bulge formation is still an open question and a topic of research \citep[e.g.,][]{Nog99,Ste02,Kor04,Dek09}. However, the phenomenological approach which we have been using so far can provide some insights here as well. \par
In order to study the $m_{\rm bh}/m_{bulge}$ relation in what follows, when inferring the bulge mass today, we compute the stellar mass which was created before  certain redshift and attribute this mass to the bulge component. This is motivated by the work of \cite{Lil16} who studied the correlation between rsSFR and internal structure of galaxies. Their simple model uses observed evolution of the size-mass relation for star-forming galaxies to model the build-up and radial dependence of stellar mass in galaxies. As galaxies grow along the Main Sequence the newly created stars are distributed in an azimuthally symmetric exponential distribution with a scale length, $h$, which has a redshift dependence of $h(z) \propto (1+z)^{-1}$. For a typical galaxy at a given redshift and mass, the outer edges follow the exponential profile, while the inner parts have profile which rises above exponential due to star formation which occurred at earlier epochs. This excess of the mass in the central parts of the galaxies can be operationally associated with the ``bulge'' component. In this model galaxies which have quenched at the earlier times have much denser cores, due to to the scale length being smaller at earlier times, and a larger fraction of these galaxies can therefore be identified as a ``bulge". \par 

\begin{figure}[ht]
	\centering
  \includegraphics[width=0.99\textwidth]{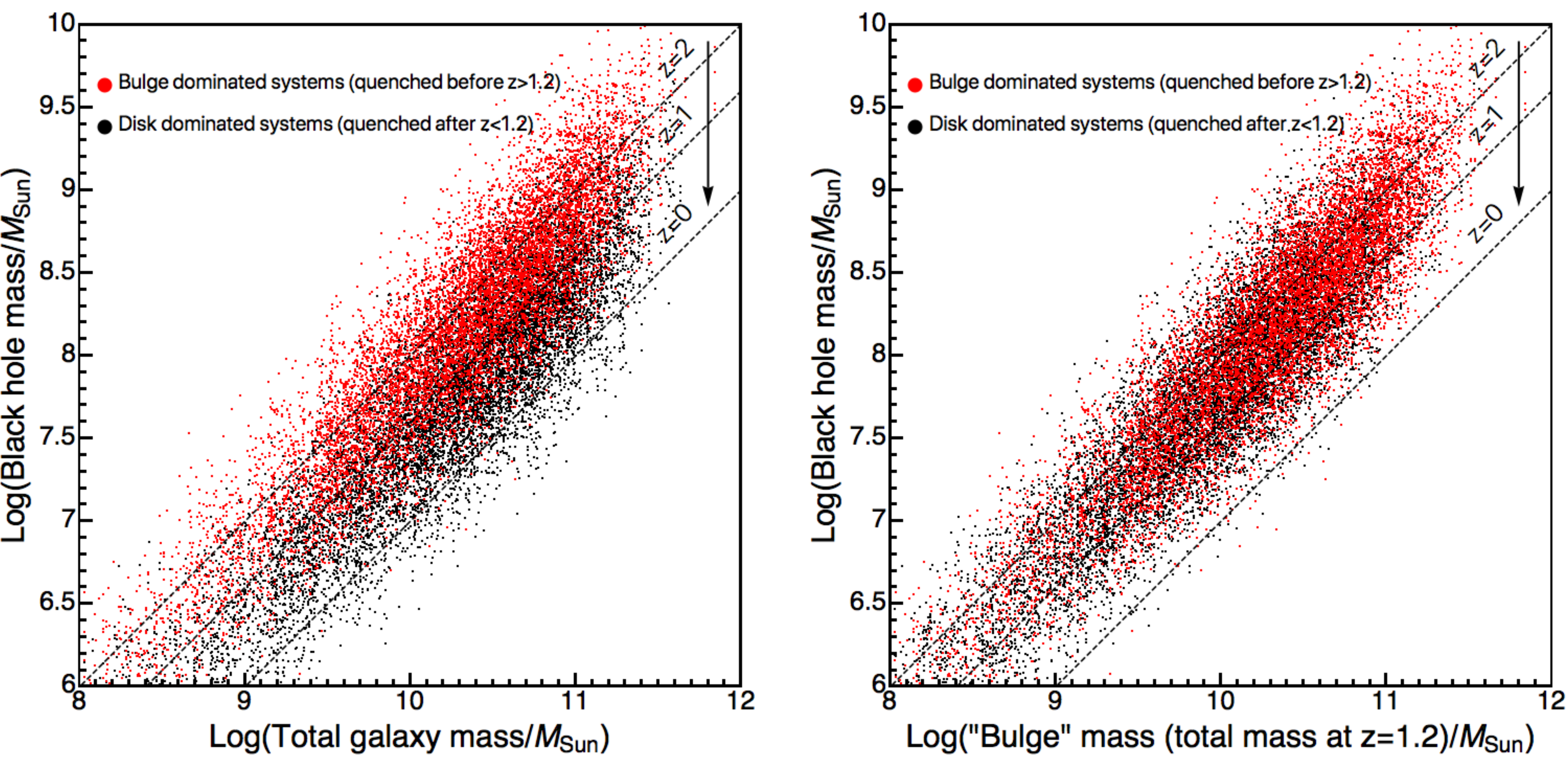}
	\caption{The correlation of black hole mass with total and bulge mass in quenched galaxies. In the left panel we show the expected $m_{\rm bh}/m_{*}$ relation in quenched galaxies. As described in the text, we simplistically separate the population of galaxies into ``bulge dominated systems", i.e., those galaxies which quenched before $z=1.2$ which are shown in red, and ``disk dominated systems", i.e., those galaxies which quenched after $z=1.2$, which are shown in black. We also show, with the dashed lines, the evolution of the typical $m_{\rm bh}/m_{*}$ ratio with redshift. In the right panel we show expected $m_{\rm bh}/m_{bulge}$ relation in quenched galaxies. In this case, the bulge mass has been simplistically computed as being the stellar mass of the galaxy at $z=1.2$.} 
	\label{fig:BulgeTotalRelation}
\end{figure}

\begin{figure}[ht]
	\centering
  \includegraphics[width=0.69\textwidth]{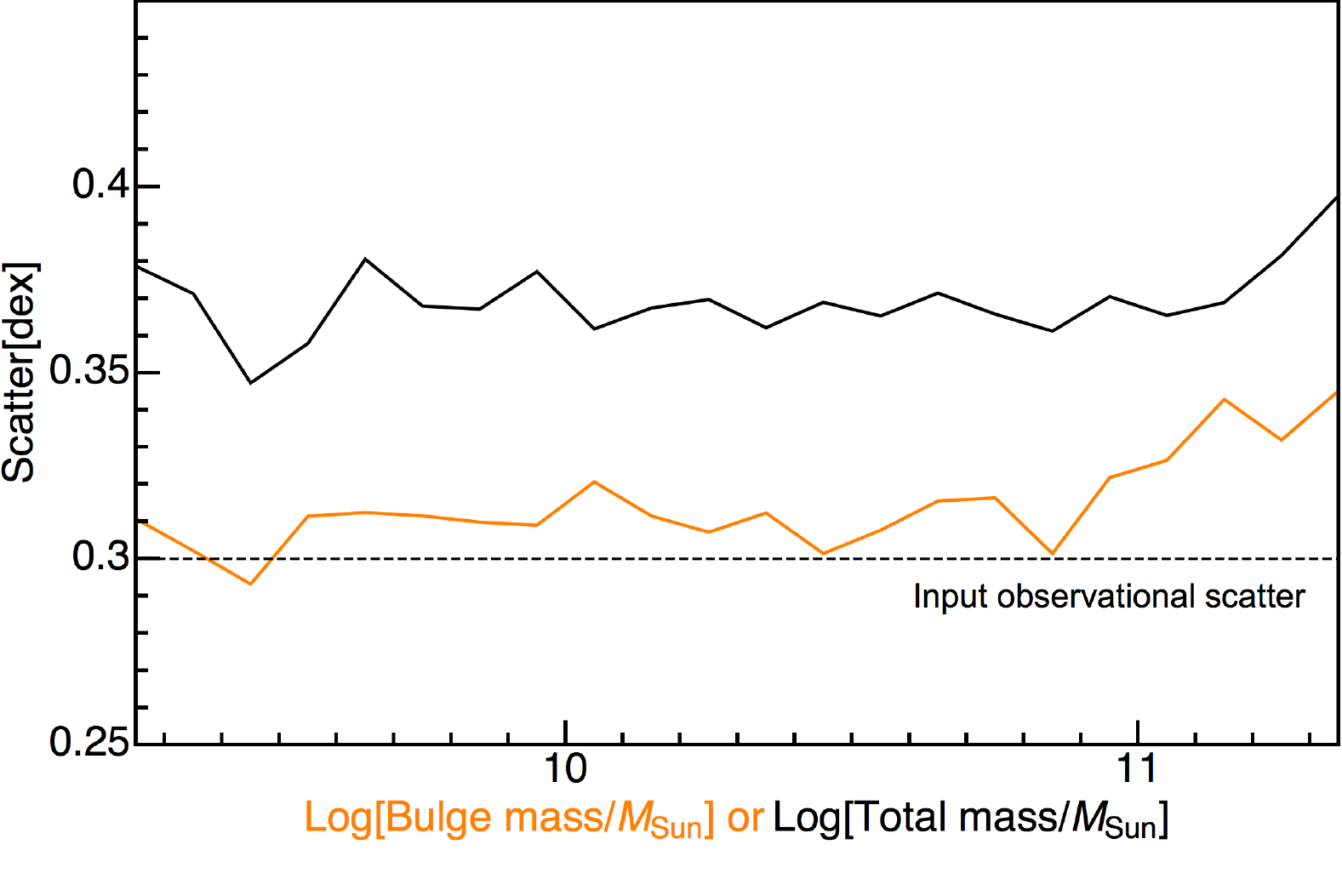}
	\caption{The scatter in the black hole - galaxy relation when computed using total mass (in black) or ``bulge" mass (in orange).  The thin dashed line shows the intrinsic observational scatter, which was set at 0.3 dex for this particular analysis.  $m_{\rm bh}/m_{bulge}$ relation shows less scatter than the $m_{\rm bh}/m_{*}$ relation. This is because $m_{\rm bh}/m_{bulge}$ relation effectively eliminates the contribution of the $m_{\rm bh}/m_{*}$ evolution to the total scatter, which is present in the $m_{\rm bh}/m_{*}$ relation. See text for discussion.} 
	\label{fig:BulgeTotalRelation2}
\end{figure}

Following the approach described above and to illustrate qualitatively the scatter expected in the $m_{\rm bh}/m_{bulge}$ relation we simply compute, for each growing galaxy, the stellar mass which was created before $z > 1.2$, and identify this as the bulge. More involved and no doubt more realistic scenarios could be surely implemented but we use this simplest assumption following the spirit of our phenomenological modeling and to transparently illustrate the main effect. We show our results in Figure \ref{fig:BulgeTotalRelation}. Comparison of the two panels clearly shows that the observed scatter is reduced when considering the $m_{\rm bh}/m_{bulge}$ relation. This is because the ``added scatter", which is caused by the redshift evolution of $m_{\rm bh}/m_{*}$ relation, is virtually eliminated when considering just the bulge parts of the galaxies, which have all been built-up at higher redshifts. $m_{\rm bh}/m_{bulge}$ relation in quenched galaxies has therefore essentially the same scatter as the intrinsic scatter of the $m_{\rm bh}/m_{*}$ relation in star-forming galaxies at a particular redshift. We show also the direct calculation of the scatter as function of either total or bulge mass in Figure \ref{fig:BulgeTotalRelation2}. For this particular calculation we used an intrinsic scatter of 0.3 dex - in this way contribution of the intrinsic scatter and the mass ratio evolution to the total observed scatter are of comparable magnitude. \par

This analysis shows how the the tight correlation between $m_{\rm bh}/m_{bulge}$ does not necessarily imply that there is deep causal connection between a black hole and its host bulge. Even though the intrinsic correlation in our model was between the black hole and the total stellar mass, the interplay between the rsSFR evolution, the galaxy size evolution, and the $m_{\rm bh}/m_{*}$ evolution can produce a tighter $m_{\rm bh}/m_{bulge}$ relation. Of course, as in the previous section, this analysis cannot recover which physical process is actually the primary driver for correlations observed in the Universe. One could also conduct the analysis ``in reverse", by arguing that the $m_{\rm bh}/m_{bulge}$ in quenched galaxies is the most fundamental correlation and drives the apparent evolution of $(m_{\rm bh}/m_{*})_{Qing}$ which we use as a premise. 

\section{Possible biases due to small size of fields with multi-wavelength coverage}
\label{sec:Small}
In this section we want to point out how the mass dependence of the $\left\langle  L_{x} \right\rangle / \left\langle SFR \right\rangle$ relation can be modified if the observational sample is lacking, for any reason, very luminous AGN. 
Due to observational constraints, these types of multi-wavelength observations are still done on relatively small fields. If the luminous AGN are absent in the sample of galaxies, this can significantly change the conclusion about dependence of mean AGN luminosity with galaxy mass because the rarest and most luminous objects are most likely to be hosted in more massive galaxies. \par 

\begin{figure*}[ht!]
    \centering
    \includegraphics[width=0.75\textwidth, height=0.60\textheight,keepaspectratio]{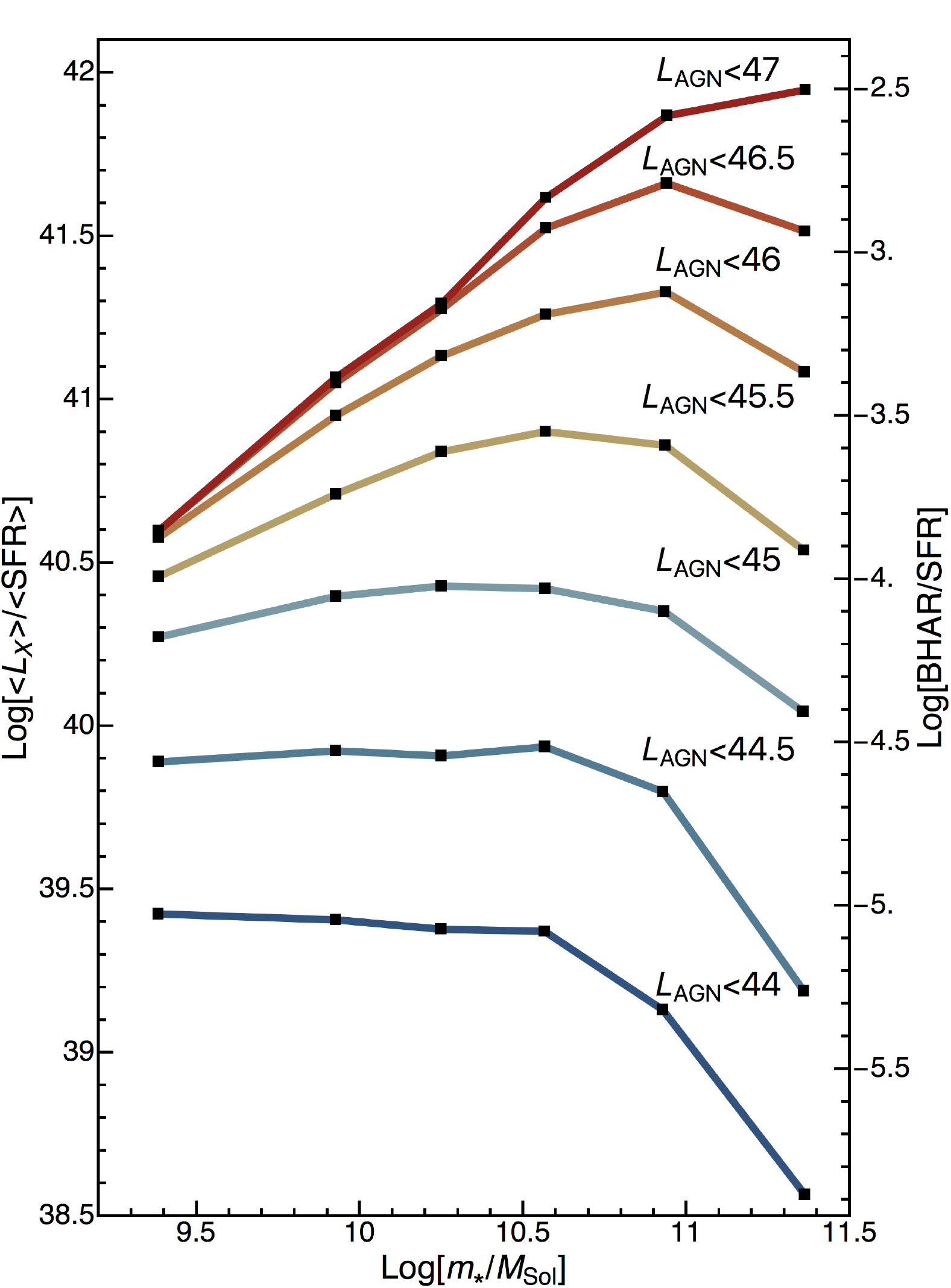}
    \caption{Schematic representation of the effect of removing AGN above a certain luminosity on the inferred $\left\langle  L_{x} \right\rangle / \left\langle SFR \right\rangle$ dependence with stellar mass. Lines show predicted $\left\langle  L_{x} \right\rangle / \left\langle SFR \right\rangle$ dependence if only AGN below a bolometric luminosity, indicated next to each line, are kept in the sample. Luminosities on the Figure are logarithmic and expressed in units of erg/s. The figure has been created assuming the survey is at $z\sim 2$, which is the time when $ L^{*} \approx 10^{46.7}$ erg/s.}
    \label{fig:EffectOfLum}
\end{figure*}

In Figure \ref{fig:EffectOfLum}, we schematically show predictions from our quenching model at redshift $z\sim 2$ with the effects we expect if more luminous objects are removed from the sample. On the x-axis, we have grouped galaxies in bins of 0.5 dex, as binning over a relatively large range of masses is commonly done in observations. On the left y-axis we denote the mean x-ray luminosity over star formation rate for galaxies in the bin. On the right-hand side, we show the equivalent quantity, black hole accretion rate over star formation rate. We use a constant offset of a factor $10^{44.45}$, as used in \cite{Rod15}. First, we consider the dark red line on the top of the figure that shows a non-biased sample. This follows the true distribution. For low masses, the slope has linear dependence, which is the consequence of the trend that the fraction of star-forming galaxies which ``are quenching'' increases linearly with the stellar mass (see Equation \eqref{eq:host}). At the high mass end there is some flattening because the fraction saturates, i.e., the fraction of star-forming galaxies that host an AGN cannot continue to grow linearly as the high enough masses fraction of star-forming galaxies that host an AGN approaches unity. 

We then start removing the most luminous objects from our sample, as indicated by the successive orange, yellow and blue lines. As we have pointed out, removing more luminous objects from the sample disproportionately affects the measurement of the mean AGN luminosities of high mass galaxies since these are more likely to host these rare very luminous AGN. This effect causes a flattening of the observed relation, because the inferred mean luminosity of AGN in high mass galaxies is underestimated. In the unrealistic extreme case in which we are only including low luminosity AGN (e.g., dark blue line), the mean measured AGN luminosity in high mass galaxies is very low, since large black holes in those galaxies generally do not radiate at these low luminosities. \par  

\end{document}